\documentclass[12pt]{article}

\usepackage{graphicx}
\usepackage{amsmath, amssymb}
  
\begin{document}
  
\begin{titlepage}

\def\thefootnote{\fnsymbol{footnote}}

\begin{center}

\hfill ICRR-Report 572-2010-5 \\
\hfill UT-HET 044 \\
\hfill UT-10-14\\
\hfill August, 2010

\vspace{0.5cm}
{\Large\bf Decaying Dark Matter in Supersymmetric Model \\
and Cosmic-Ray Observations}

\vspace{1cm}
{\large Koji Ishiwata}$^{\it (a)}$, 
{\large Shigeki Matsumoto}$^{\it (b)}$, 
{\large Takeo Moroi}$^{\it (c,d)}$

\vspace{1cm}

{\it $^{(a)}${Institute for Cosmic Ray Research, University of Tokyo, \\
    Kashiwa 277-8582, Japan}}

\vspace{0.2cm}

{\it $^{(b)}${Department of Physics, University of Toyama, 
    Toyama 930-8555, Japan}}

\vspace{0.2cm}

{\it $^{(c)}${Department of Physics, University of Tokyo, Tokyo
    113-0033, Japan}}

\vspace{0.2cm}

{\it $^{(d)}${Institute for the Physics and Mathematics of the
    Universe, University of Tokyo, \\ Kashiwa 277-8583, Japan}}

\vspace{1cm}

\abstract{We study cosmic-rays in decaying dark matter scenario,
  assuming that the dark matter is the lightest superparticle and it
  decays through a $R$-parity violating operator. We calculate the
  fluxes of cosmic-rays from the decay of the dark matter and those
  from the standard astrophysical phenomena in the same propagation
  model using the GALPROP package. We reevaluate the preferred
  parameters characterizing standard astrophysical cosmic-ray sources
  with taking account of the effects of dark matter decay. We show
  that, if energetic leptons are produced by the decay of the dark
  matter, the fluxes of cosmic-ray positron and electron can be in
  good agreements with both PAMELA and Fermi-LAT data in wide
  parameter region. It is also discussed that, in the case where
  sizable number of hadrons are also produced by the decay of the dark
  matter, the mass of the dark matter is constrained to be less than
  200$-$300 GeV in order to avoid the overproduction of
  anti-proton. We also show that the cosmic $\gamma$-ray flux can be
  consistent with the results of Fermi-LAT observation if the mass of
  the dark matter is smaller than $\sim 4~{\rm TeV}$.}

\end{center}
\end{titlepage}

\renewcommand{\theequation}{\thesection.\arabic{equation}}
\renewcommand{\thepage}{\arabic{page}}
\setcounter{page}{1}
\renewcommand{\thefootnote}{\#\arabic{footnote}}
\setcounter{footnote}{0}

\section{Introduction}
\label{sec:intro}
\setcounter{equation}{0}

An anomaly has been discovered in the recent observation of cosmic-ray
positron. The PAMELA satellite has observed positron excess in the
flux around the energy range of 10$-$100 GeV, which cannot be
explained by standard astrophysical phenomena in our
Galaxy~\cite{Adriani:2008zr}. A possible solution is to consider
non-standard phenomena in astrophysics such as nearby
pulsars~\cite{Hooper:2008kg}. On the other hand, the PAMELA experiment
has motivated us to settle this problem in the viewpoint of particle
physics; the anomaly can be explained by the decay or the annihilation
of dark matter of the universe. Indeed, various scenarios with
decaying~\cite{Nomura:2008ru}$-$\cite{Ishiwata:2008cv} and
annihilating~\cite{Cirelli:2008pk}$-$\cite{Harnik:2008uu} dark matter
have been discussed in much literature.\footnote{For the decaying
  dark matter scenario, it was pointed out that unstable dark matter
  seems to be suggested by the HEAT results~\cite{DuVernois:2001bb}
  and that the signal of decaying dark matter could be observed by
  PAMELA as a positron excess before the PAMELA result showed
  up~\cite{Ishiwata:2008cu,Ibarra:2008qg}.}

A well-motivated scenarios is decaying dark matter scenario in the
framework of supersymmetric model. Supersymmetric model provides
natural candidate for dark matter, {\it i.e.}, the lightest
superparticle (LSP). In addition, supersymmetric model is one of the
most attractive models beyond the standard model; it gives a solution
to hierarchy problem and enables the three gauge coupling constants to
unify at $\sim 10^{16}\ {\rm GeV}$. A very long lifetime of the LSP,
which is required to solve the PAMELA anomaly, can be realized with a
very weak $R$-parity violation (RPV). In the scenario with unstable
LSP dark matter, not only positron but also electron, $\gamma$-ray,
and/or anti-proton give significant contribution to cosmic-ray fluxes.
Thus, in order to solve the PAMELA anomaly in this framework, it is
necessary to check if the fluxes of all the cosmic-ray particles are
in agreement with observation. For such a purpose, one should
calculate the fluxes of cosmic-rays from standard astrophysical
source, {\it i.e.}, supernova remnant (SNR), as well as those from the
decay of dark matter. In many of the previous studies, however, fluxes
of cosmic-rays from different sources are evaluated using different
propagation models and/or different SNR injection parameters. For a
complete study of the decaying dark matter scenario, the fluxes of all
the cosmic-ray species should be calculated in a consistent way.

In this paper, we consider the case that dark matter is unstable (and
long-lived) and it becomes the significant source of high energy
cosmic-rays. We pay particular attention to the case where the LSP is
dark matter, and study the cosmic-ray fluxes in light of PAMELA and
other cosmic-ray observations. We use the GALPROP
package~\cite{Galprop} so that the cosmic-ray fluxes from different
sources are evaluated with the same propagation models.\footnote{For
  the former study of decaying scenario by using GALPROP, see
  Refs~\cite{Yin:2008bs}, \cite{Zhang:2008tb}$-$\cite{Cotta:2010ej}.}
We reevaluate some of the parameters characterizing the SNR injection
spectra taking account of the effects of the decay of dark matter. We
then discuss constraints on the decaying scenario from
observations. We will show that, by properly choosing model
parameters, the PAMELA anomaly can be well explained in the decaying
dark matter scenario without conflicting other constraints.

The organization of this article is as follows. In the next section,
we explain how we calculate cosmic-rays from the standard
astrophysical phenomena and decaying dark matter. Here important
parameters in our analysis are described. Numerical results are then
shown in section~\ref{sec:results}. Section~\ref{sec:conclusion} is
devoted to conclusion.

\section{Cosmic-Rays from SNR and Dark Matter}
\label{sec:CRsimulation}
\setcounter{equation}{0}

Spectra of cosmic-ray particles which we observe depend on what the
sources of cosmic-rays are and how the particles from the sources
propagate. In this section, we first consider cosmic-rays from SNR
which is supposed to be standard source of high energy
cosmic-rays. Cosmic-rays from decaying dark matter are then discussed.
In our numerical calculation, all the cosmic-ray fluxes are obtained
by using the single numerical code, GALPROP, in order to perform a
consistent analysis.  We explain relevant parameters characterizing
energetic cosmic-rays from astrophysical sources and dark matter
decay.

\subsection{Cosmic-rays in standard Galactic model}
\label{sec:BG}

Cosmic-rays mainly consist of nuclei, electron, positron, anti-proton,
and $\gamma$-ray. Among those, electron and nuclei, {\it e.g.},
proton, helium, carbon, oxygen, and iron, are considered to be from
the remnants of supernovae and pouring to the earth after they have
drifted by interaction with interstellar matters and magnetic field in
our Galaxy. These are called primary cosmic-rays. On the other hand,
cosmic-rays are also generated secondarily in a consequence of
interaction processes; we call them secondary cosmic-rays. (See
Table~\ref{table:CR_gal}.) In the collisions of cosmic-ray with
interstellar gas, such as $pp$-collision, anti-protons and other
nuclei such as lithium, beryllium, boron, and sub-Fe (scandium,
titanium, and vanadium), are produced. In the collision processes,
pion and kaon are also produced; secondary positron and electron are
then emitted in the cascade decay of such mesons.

\begin{table}[t]
 \begin{center}
  \begin{tabular}{llc}
   \hline \hline
   Source  & (Propagation and interaction process) & Products \\
   \hline
   SNR $e^-$  & $\rightarrow$ (propagate) $\rightarrow$ & $e^-_{\rm prim}$ \\
   SNR $p$  & $\rightarrow$ ($pp$-collision) $\rightarrow$ ($\pi,K$ decay) 
   $\rightarrow$   & $e^{\pm}_{\rm sec}$ \\ 
   \hline
   SNR $p$  & $\rightarrow$ (propagate) $\rightarrow$ & $p_{\rm prim}$ \\
   SNR $p$  & $\rightarrow$ ($pp$-collision) $\rightarrow$   
   & $p_{\rm sec},\bar{p}_{\rm sec}$ \\
   \hline
   $e^-_{\rm prim/sec}$, $e^+_{\rm sec}$  
   & $\rightarrow$  (IC + bremss + synch) $\rightarrow$& 
   $\gamma$ \\
   SNR $p$  & $\rightarrow$ ($pp$-collision) $\rightarrow$ ($\pi,K$ decay) 
   $\rightarrow$   & $\gamma$ \\  
   \hline \hline
  \end{tabular}
  \caption{\small Propagation and production of cosmic-ray $e^{\pm}$,
    $p,~\bar{p}$, and $\gamma$ from SNR. Here $e^-_{\rm prim}$
    ($p_{\rm prim}$) and $e^{\pm}_{\rm sec}$ ($p_{\rm sec}$,
    $\bar{p}_{\rm sec}$) are primary and secondary $e^-$ ($p$) and
    $e^{\pm}$ ($p$, $\bar{p}$), respectively.}
  \label{table:CR_gal}
 \end{center}
\end{table}

The primary spectra of electron and proton from SNR are assumed to
obey power law, since they are assumed to be produced through
Fermi-acceleration mechanism. The source terms of these cosmic-ray
particles are then parametrized as
\begin{eqnarray}
 Q^e_{\rm SNR} &=&  A_{e^-} E_{{\rm GeV}}^{-\gamma^e},
\\
 Q^p_{\rm SNR} &=& A_{p} p_{{\rm GeV}}^{-\gamma^p}.
\end{eqnarray}
Here, $A_{e^-}$ and $\gamma^e$ ($A_p$ and $\gamma^p$) are
normalization and power index of electron (proton) injected from SNR,
while $E_{\rm GeV}$ ($p_{\rm GeV}$) is GeV-normalized energy
(momentum) of electron (proton). Following the treatment adopted in
``Conventional model'' of
cosmic-ray~\cite{Strong:1998fr,Strong:2004de}, the power indices are
assumed to take different values for high and low energy region as in
Table~\ref{table:galmodel}. If SNR is the dominant source of high
energy cosmic-rays, the shapes of those spectra are sensitive to the
index parameters $\gamma^e$ and $\gamma^p$. In particular, primary
electron generally dominates the total ($e^++e^-$) flux, and its
spectrum in the energy region over 10 GeV is sensitive to the value of
$\gamma^e$.

If we adopt the Conventional model, predicted cosmic-ray fluxes such
as boron to carbon (B/C) ratio, proton, helium, and anti-proton fluxes
are in good agreements with observations. The values of parameters
adopted in the model are summarized in
Table~\ref{table:galmodel}.\footnote{One can consider diffusive
  reacceleration or diffusive convection; we adopt the former one.}
Within this model, however, cosmic-ray $e^+$ from SNR cannot explain
the positron fraction data reported by PAMELA, which is now known as
the ``PAMELA anomaly''. In addition, the ($e^+ + e^-$) spectrum
observed by Fermi-LAT~\cite{Abdo:2009zk} is slightly harder than the
prediction of the model; however, as we will discuss later, such a
discrepancy may be solved by adopting an appropriate value of
$\gamma^e$.

In the study of high energy cosmic-ray, it is also important to
consider cosmic $\gamma$-ray and radiation fluxes. Inside the Galaxy,
high energy $\gamma$-ray is necessarily produced by several processes
(see Table~\ref{table:CR_gal}): the inverse-Compton (IC) scattering of
$e^{\pm}$ with interstellar radiation field (ISRF), the bremsstrahlung
of $e^{\pm}$ with interstellar gas, the synchrotron radiation from
$e^{\pm}$ under magnetic field, and the decay of pion which is
produced in the hadronic reaction of cosmic nuclei in interstellar
gas. The spectra of $\gamma$-ray and synchrotron radiation depend on
the injection spectra of SNR electron and proton. According to
Ref.~\cite{Strong:1998fr}, the Conventional model gives $\gamma$-ray
flux well below the observed flux and radiation flux consistent with
observation. It is also shown that the consistency is hold if
$\gamma^e \lesssim 2.0$ for $E \lesssim 10$ GeV is satisfied
\cite{Strong:1998fr}.\footnote
{In order to be consistent with local $e^-$ flux in $1~{\rm GeV} \le E
  \le 30~{\rm GeV}$, the power index of injected electron $\gamma^e
  \sim 2.5-2.7$ for $E\gtrsim O(1~{\rm GeV})$ is favored.}
Electron in this energy range is, however, considered to be subject to
relatively large effect of the solar modulation.

\begin{table}[t]
  \begin{center}
    \begin{tabular}{lccc}
      \hline \hline
      $\Phi_{e^-_{\rm prim}}^{(0)}$ 
      &1.3$\times 10^{-11}$ 
      ${\rm GeV}^{-1}{\rm m}^{-2}{\rm sec}^{-1}{\rm str}^{-1}$ (at 100 GeV)\\
    $\gamma^e$   &2.54/1.6  (above/below 4 GeV)\\
    \hline
    $\Phi_{p_{\rm prim}}^{(0)}$ 
    &5.0$\times 10^{-2}$ 
    ${\rm GeV}^{-1}{\rm m}^{-2}{\rm sec}^{-1}{\rm str}^{-1}$ (at 100 GeV)\\
    $\gamma^p$   & 2.42/1.98  (above/below 9 GeV)\\
    \hline \hline
    \end{tabular}
    \caption{\small Important parameters of cosmic-ray used to
      calculate fluxes from SNR. The primary electron (proton) is
      normalized to fit observations, and $\gamma^e$ ($\gamma^p$) is
      chosen to reproduce the observed spectrum. Those values are
      adopted in Conventional model.}
      \label{table:galmodel}
  \end{center}
\end{table}

Since we are interested in the scenario where cosmic-rays are produced
from the decay of dark matter, we do not necessarily restrict
ourselves to the Conventional model. We should rather reevaluate some
of the parameters characterizing the SNR injection spectra with taking
account of the effects of the decaying dark matter.  In this paper, we
take $A_{e^-}$ and $\gamma^e$ to be free parameters. (The other
parameters are taken as the same as in Conventional model unless
otherwise mentioned.)

Hereafter, the cosmic-rays originating in SNR are called as background (BG):
\begin{eqnarray}
  \Phi_{e^-_{\rm BG}} &=& \Phi_{e^-_{\rm prim}} + \Phi_{e^-_{\rm sec}},
  \\
  \Phi_{e^+_{\rm BG}} &=& \Phi_{e^+_{\rm sec}},
\end{eqnarray}
where $\Phi_{e^-_{\rm prim}}$ and $ \Phi_{e^{\pm}_{\rm sec}}$ are
fluxes of primary $e^-$ and secondary $e^{\pm}$, respectively. Besides
electron and positron, hadrons are also produced which result in
background proton and anti-proton. We also calculate BG fluxes of
these particles:
\begin{eqnarray}
  \Phi_{p_{\rm BG}} &=& \Phi_{p_{\rm prim}} + \Phi_{p_{\rm sec}},
  \\ 
  \Phi_{{\bar p}_{\rm BG}} &=& \Phi_{{\bar p}_{\rm sec}},
\end{eqnarray}
where $\Phi_{p_{\rm prim}}$ is primary $p$ flux, while $\Phi_{p_{\rm
    sec}}$ ($\Phi_{{\bar p}_{\rm sec}}$) is secondary $p$ ($\bar{p}$)
flux. In the calculation of cosmic-ray fluxes of $p$ and $\bar{p}$, we
use force-field model~\cite{solarmd} with the use of the solar
modulation potential of $\phi=550$ MV.\footnote{This is a method
  proposed in order to take into account the effect of solar
  modulation. Estimation of the effect of solar modulation is,
  however, still uncertain. We thus focus on the energy region where
  the effect is small enough in our numerical analysis. (See the
  discussion below.)}  In addition, for later convenience, we
introduce normalization parameter (donated as $a_e$), which is defined
by $\Phi_{e^-_{\rm prim}}(E) = a_e \Phi_{e^-_{\rm prim}}^{(0)}(E)$,
with $\Phi_{e^-_{\rm prim}}^{(0)}$ being the reference flux normalized
as $\Phi_{e^-_{\rm prim}}^{(0)} (E = 100\ {\rm GeV}) = 1.3 \times
10^{-11}~{\rm GeV}^{-1}{\rm m}^{-2}{\rm sec}^{-1}{\rm str}^{-1}$.

For the estimation of the BG $\gamma$-ray flux, we also use the
GALPROP package. There may exist other astrophysical contributions
which are not taken into account in the GALPROP package; we do not
consider such contributions in our analysis because they are expected
to have large uncertainties.\footnote{Unidentified cosmic $\gamma$-ray
  may have various origins, for example, galaxy
  clusters~\cite{Ensslin:1996ep}, energetic particles in the shock
  waves associated with large-scale cosmological structure
  formation~\cite{Loeb:2000na}, distant gamma-ray burst events,
  baryon-antibaryon annihilation~\cite{Gao:1990bh}. }

\subsection{Cosmic-rays from decaying dark matter}
\label{sec:CRDM}

Now we consider cosmic-rays from the decay of dark matter. In the
decaying scenario, it is assumed that the lifetime of dark matter is
much longer than the present age of the universe. Most of dark matter
therefore survives today. Even though the decay of dark matter is
suppressed by the long lifetime, it can be a significant source of
cosmic-ray if standard-model particles are produced through the
decay. Cosmic-rays from dark matter depend on the spectra of particles
injected from the decay and how they propagate in the universe.

Important quantities in the calculation of comic-ray fluxes from
unstable dark matter are lifetime of dark matter ($\tau_{\rm DM}$),
mass of dark matter ($m_{\rm DM}$), and spectra of emitted
particles. With these, source term of cosmic-ray is given by
\begin{eqnarray}
  Q_{\rm DM}^X = 
  \frac{1}{\tau_{\rm DM}}\frac{\rho_{\rm DM}}{m_{\rm DM}} \frac{dN^X}{dE},
\end{eqnarray} 
where $X$ is cosmic-ray particle ({\it e.g.}, $X=e^{\pm}, p,\bar{p}$,
and $\gamma$), and $dN^X/dE$ is the spectrum of $X$ from the decay of
a single dark matter. In addition, $\rho_{\rm DM}$ is mass density of
dark matter. In the decaying scenario, the fluxes of cosmic-rays are
insensitive to the profile of dark matter density except for comic-ray
$\gamma$ from Galactic center.  Since we will not study such
$\gamma$-ray, we adopt the isothermal profile for Galactic halo;
\begin{eqnarray}
  \rho^{\rm (Galaxy)}_{\rm DM} (r) 
  =
  \rho_\odot \frac{r_{\rm core}^2 + r_\odot^2}{r_{\rm core}^2+r^2},
  \label{eq:isothermal}
\end{eqnarray}
where $\rho_\odot\simeq 0.43\ {\rm GeV/cm^3}$ is the local halo
density, $r_{\rm core}\simeq 2.8\ {\rm kpc}$ is the core radius,
$r_\odot\simeq 8.5\ {\rm kpc}$ is the distance between the Galactic
center and the solar system, and $r$ is the distance from the Galactic
center. We have checked that our numerical results are almost
unchanged even if we use other dark matter profiles such as the NFW
profile~\cite{Navarro:1996gj}. 

In the decaying dark matter scenario, cosmic $\gamma$-ray from
extra-Galactic region may have significant flux because the
$\gamma$-ray is produced by the IC process induced by the energetic
$e^\pm$ from dark matter decay \cite{Ishiwata:2009dk}.  It was shown
that the flux could be comparable to the observed $\gamma$-ray
flux. (For the discussion of constraint from isotropic $\gamma$-ray
observation, see also \cite{Cirelli:2009dv,Hutsi:2010ai}.)  We
calculate the extra-Galactic $\gamma$-ray by following formula given
in \cite{Ishiwata:2009dk}.  In the calculation of cosmic $\gamma$-ray
from extra-Galactic region, we use averaged dark matter density
as\footnote
{Because we only consider the $\gamma$-ray flux averaged over large
  solid angle in the decaying dark matter scenario, we do not have to
  worry about the effects of sub-halo. These effects may be important
  in the annihilating dark matter scenario as shown in
  Refs.~\cite{Cline:2010ag}$-$\cite{Perelstein:2010at}.}
\begin{eqnarray}
  \rho_{\rm DM}^{\rm (extra\text{-}Galaxy)} = 
  1.2 \times 10^{-6}~{\rm GeV/cm}^3.
\end{eqnarray}
Notice that the $\gamma$-ray from the extra-Galactic region is
isotropic.  $\gamma$-ray is also produced at the central region of
our Galaxy.  However, the flux from the Galactic center strongly
depend on the dark matter profile \cite{Cirelli:2009dv}.  Thus, in
order to derive conservative constraint, we only consider $\gamma$-ray
from high Galactic latitude.

\begin{table}[t]
 \begin{center}
  \begin{tabular}{lcccc}
    \hline \hline
    Final state  & &  Primary & & Secondary \\
    \hline
    Leptons &  $\rightarrow$
    & $e^{\pm}$ &
    $\rightarrow$  & $\gamma_{\rm IC}$ \\ 
    Leptons + hadrons&  $\rightarrow$
    & $e^{\pm},~p,~\bar{p},~\gamma$ &
    $\rightarrow$  & $\gamma_{\rm IC}$ \\
    \hline \hline
  \end{tabular}
  \caption{\small Production process of cosmic-rays in the decaying
    dark matter scenario we consider. Here, $\gamma_{\rm IC}$ means
    $\gamma$-ray produced in inverse-Compton scattering process.}
  \label{table:CR_DM}
 \end{center}
\end{table}

In supersymmetric model, there are many candidates for dark matter
which can decays under RPV. One of the possibilities motivated by the
PAMELA anomaly is unstable dark matter which decays into final state
which consists of only energetic leptons. Then, charged leptons can
make prominent rise in positron fraction to explain the anomaly
without conflicting the observation of cosmic-ray anti-proton. In such
a case, $\gamma$-ray is inevitably produced by IC process as secondary
cosmic-ray. (Those production mechanism are summarized in
Table~\ref{table:CR_DM}, including the following case to be described
below.) On the other hand, one can also consider a case where
significant amount of hadrons (as well as leptons) are contained in
the final state of the decay of dark matter. In such a case,
anti-proton as well as $\gamma$-ray may provide stringent constraints
on the scenario. Here, notice that scenario with similar final state
results in similar cosmic-ray fluxes. We thus focus on several cases
which give typical results; gravitino and sneutrino dark matter
scenarios with bi-linear and/or tri-linear RPV. In each case, we
calculate $dN^X/dE$, including contributions of particles form
cascading decay after hadronization, with the use of the PYTHIA
package~\cite{Sjostrand:2006za}.

Let us first consider the case where gravitino is the LSP.  In the
present case, because of the smallness of RPV, the production
mechanism of gravitino is unaffected by the RPV interaction; gravitino
can be produced by scattering processes of thermal particles
\cite{Moroi:1993mb}, decay of the LSP in minimal supersymmetric
standard model sector (MSSM-LSP) after freeze-out \cite{Moroi:1993mb,
  Feng:2003xh} or a non-thermal process
\cite{Ishiwata:2007bt}.\footnote
{However, the production via the decay of the MSSM-LSP is suppressed
  if the lifetime of the MSSM-LSP becomes shorter than the cosmic time
  of its freeze-out.}
(Sneutrino, which we will discuss below, can be also produced by
non-thermal process \cite{Asaka:2005cn}).  The spectra of final-state
particles, on the other hand, strongly depend on how the $R$-parity is
violated. Because we are interested in the PAMELA anomaly, we only
consider RPV operators by which the LSP decays into final states with
energetic lepton. Then, one possibility is to introduce the following
soft breaking bi-linear RPV operators:
\begin{eqnarray}
  {\cal L}_{\rm bi {\text -} RPV} 
  = B_i \tilde{L}_i H_u + m^2_{\tilde{L}_i H_d} \tilde{L}_i H^*_d 
  + {\rm h.c.},
  \label{eq:biRPV}
\end{eqnarray}
where $\tilde{L}_i$ is left-handed slepton doublet (with $i=1-3$ being
the generation index), while $H_u$ and $H_d$ are up- and down-type
Higgs boson doublets, respectively. (Here, $L_i$ and $H_d$ are defined
in the bases in which the bi-linear RPV terms in superpotential are
rotated away.) With the operators given in Eq.~\eqref{eq:biRPV},
gravitino decays into two-body final states: $\psi_{\mu} \rightarrow
\gamma \nu_i$, $Z\nu_i$, $Wl_i$, and $h\nu_i$. (Here and thereafter,
gravitino is denoted as $\psi_\mu$.) Decay width of each process is
given in Ref.~\cite{Ishiwata:2008cu}; the decay process is dominated
by the mode $\psi_{\mu} \rightarrow Wl_i$ if the gravitino is heavier
than $W$-boson.  The lifetime is related to the bi-linear RPV
parameter as \cite{Ishiwata:2008cu},
\begin{eqnarray}
  \tau_{3/2}\simeq 8 \times 10^{25}\ {\rm sec} \times
  \left( \frac{\kappa}{10^{-8}} \right)^{-2} 
  \left( \frac{m_{3/2}}{200\ {\rm GeV}} \right)^{-3},
\end{eqnarray}
where $m_{3/2}$ is gravitino mass.  In addition, $\kappa^2\equiv\sum_i
\kappa_i^2$, where
\begin{eqnarray}
  \kappa_i = \frac{B_i \sin\beta + m^2_{\tilde{L}_i H_d} \cos\beta}
  {m^2_{\tilde{\nu}_{i}}},
\end{eqnarray}
with $\tan\beta = \langle H^0_u \rangle / \langle H^0_d \rangle$ (here
$\langle H^0_u \rangle$ and $\langle H^0_d \rangle$ are the vacuum
expectation values of up- and down-type Higgs bosons, respectively),
and $m_{\tilde{\nu}_i}$ being the mass of sneutrino in $i$-th
generation.  We will see that the lifetime of dark matter should be
$O(10^{26}~{\rm sec})$ in order to explain the PAMELA anomaly.  Such a
lifetime is realized when $\kappa \sim O(10^{-9}-10^{-10})$.

Another important possibility is the decay through tri-linear RPV:
\begin{eqnarray}
  W_{\rm tri {\text -} RPV}
  =
  \frac{1}{2} \lambda_{ijk}\hat{L}_i \hat{L}_j \hat{E}^c_k,
\label{eq:W_LLE}
\end{eqnarray}
where ${\hat L}_i=({\hat \nu}_{Li}, {\hat l}_{Li})$ and $\hat{E}^c_i$
are left-handed lepton doublet and right-handed lepton singlet,
respectively, where ``hat'' is for superfield. With the operator given
in Eq.\ \eqref{eq:W_LLE}, gravitino decays as $\psi_{\mu}\rightarrow
\nu_i l^{\pm}_{L,j}l^{\mp}_{R,k}$ through diagrams with virtual
slepton. Here, for simplicity, we assume that the right-handed
sleptons are much lighter than the left-handed ones. The energy
distribution of the final-state leptons is given
\begin{eqnarray}
  \frac{d \Gamma_{\psi_{\mu}\rightarrow \nu l_{L} l_{R}}}
  {d E_{l_L}d E_{l_R}} &=& \frac{\lambda^2 m_{3/2}}{768\pi^3 M_{\rm Pl}^2}
  \frac{z_R^3 (1-z_R)}{(m_{\tilde{l}_R}/m_{3/2})^2 - (1-z_R)},
\end{eqnarray}
where $M_{\rm Pl}\simeq 2.4\times 10^{18}\ {\rm GeV}$ is the reduced
Planck scale, $m_{\tilde{l}_R}$ is the right-handed slepton mass, and
$z_R = 2E_{l_R}/m_{3/2}$. In our analysis, $\lambda_{ijk}$ is
determined so that the preferred value of the lifetime of gravitino is
obtained.  Notice that, taking $m_{\tilde{l}_R}\sim m_{3/2}$ for
simplicity, the lifetime is estimated as
\begin{eqnarray}
\tau_{3/2}\simeq 7 \times 10^{26}~{\rm sec} \times
\left( \frac{\lambda}{10^{-8}}\right)^{-2} 
\left( \frac{m_{3/2}}{200~{\rm GeV}}\right)^{-3}.
\end{eqnarray}
Here, $\lambda_{ijk}$ must satisfy the condition $\lambda_{ijk}
\lesssim 10^{-7}$ in order not to wash out baryon
asymmetry~\cite{Campbell:1990fa}; this constraint is satisfied in the
parameter space we will be interested in.

If the sneutrino is the LSP, monochromatic leptons are produced by its
decay via the tri-linear RPV superpotential given in
Eq.~\eqref{eq:W_LLE}. Indeed, the sneutrino can decay as
$\tilde{\nu}\rightarrow l^+_L l^-_R$. Because the sneutrino is a
viable candidate for dark matter irrespective of its handedness
\cite{Hall:1997ah, Asaka:2005cn}, and also because the fluxes of high
energy cosmic-rays are sensitive to the spectra of emitted particles
from the decay of dark matter, we also consider the case of unstable
sneutrino LSP.  In this case, the lifetime is determined as
\cite{Ishiwata:2009vx},
\begin{eqnarray}
\tau_{\tilde{\nu}} \sim
2\times 10^{26}~{\rm sec}
\left(\frac{\lambda \sin\theta_{\tilde{\nu}}}{10^{-26}}\right)^{-2}
\left(\frac{m_{\tilde{\nu}}}{200~{\rm GeV}}\right)^{-1},
\end{eqnarray}
where $\theta_{\tilde{\nu}}$ is the sneutrino mixing angle.

\section{Numerical Results}
\label{sec:results}
\setcounter{equation}{0}

\begin{table}[t]
 \begin{center}
  {\footnotesize
  \begin{tabular}{c|cccccccc}
   & Decay mode & $m_{\rm DM}$ (TeV) & $\tau_{\rm DM}$ (sec) & $a_e$ & $\gamma^e$
   & $\gamma^p$ & $\chi^2_{\rm Pamela}$ & $\chi^2_e$ \\
   \hline
   \hline
   (I-1)
   & $\psi_{\mu} \to eW, \nu_e Z,$
   & 0.20 & 9.6 $\times$ 10$^{26}$ & 1.02 & 2.35 & 2.42 & 9.8 & 41.0 \\
   (I-2)
   & \qquad~~$\nu_e h, \nu_e \gamma$
   & 1.0 & 2.5 $\times$ 10$^{26}$ & 0.80 & 2.60 & 2.42 & 11.0 & 69.6 \\
   (I-3)
   & & 4.0 & 1.7 $\times$ 10$^{26}$ & 0.88 & 2.54 & 2.42 & 8.8 & 20.5 \\
   (I-4)
   & & 0.20 & 8.4 $\times$ 10$^{26}$ & 1.02 & 2.34 & 2.52 & 9.8 & 46.3 \\
   (I-5)
   & $\psi_{\mu} \to \mu W, \nu_\mu Z,$
   & 0.25 & 4.2 $\times$ 10$^{26}$ & 1.02 & 2.42 & 2.42 & 10.9 & 43.6 \\
   (I-6)
   & \qquad~~$\nu_\mu h, \nu_\mu \gamma$
   & 1.0 & 2.6 $\times$ 10$^{26}$ & 0.94 & 2.43 & 2.42 & 10.7 & 45.8 \\
   (I-7)
   & & 4.0 & 1.5 $\times$ 10$^{26}$ & 0.88 & 2.52 & 2.42 & 7.2 & 10.0 \\
   (I-8)
   & & 0.25 & 3.7 $\times$ 10$^{26}$ & 1.04 & 2.41 & 2.52 & 10.9 & 46.4 \\
   \hline
   (II-1)
   & $\psi_{\mu} \to \nu \mu \tau$ & 0.25 & 9.2 $\times$ 10$^{26}$
   & 1.02 & 2.36 & 2.42 & 10.5 & 31.8 \\
   (II-2)
   & & 1.0 &4.4 $\times$ 10$^{26}$ & 0.90 & 2.46 & 2.42 & 11.0 & 49.4 \\
   (II-3)
   & & 4.0 & 1.4 $\times$ 10$^{26}$ & 0.72 & 2.62 & 2.42 & 11.0 & 98.8 \\
   \hline
   (III-1)
   & $\tilde{\nu} \to \mu\mu$
   & 0.40 & 9.6 $\times$ 10$^{26}$ & 1.00 & 2.36 & 2.42 & 6.0 & 42.1 \\
   (III-2)
   & & 1.0 & 5.7 $\times$ 10$^{26}$ & 0.90 & 2.44 & 2.42 & 10.9 & 75.8 \\
   (III-3)
   & & 4.0 & 1.8 $\times$ 10$^{26}$ & 0.68 & 2.58 & 2.42 & 10.9 & 250.6 \\
   \hline
  \end{tabular}
  }
  \caption{\small Scenarios of decaying dark matter considered in this
    paper. Model parameters giving the ``best-fit'' results to
    cosmic-ray observations are also shown in each case.}
  \label{table:Summary}
 \end{center}
\end{table}

Now we are at the position to quantitatively discuss the cosmic-ray
fluxes. As we have already mentioned in the previous section, we
assume that the dominant sources of the cosmic-rays are supernova
remnants and decaying dark matter. In order to study the cosmic-ray
fluxes from both sources using the same propagation model as well as
to take account of the production of secondary cosmic-rays, we utilize
the GALPROP package for the calculation of all the cosmic-ray fluxes
(except for the extra-Galactic $\gamma$-ray flux).

The spectral shape of each cosmic-ray particle from dark matter
depends on properties of the decay of dark matter. We take $m_{\rm
  DM}$ and $\tau_{\rm DM}$ as free parameters in this paper, and
consider three typical cases summarized in Table.~\ref{table:Summary};
the cases with (I) the emission of a charged lepton as well as a weak
boson resulting in energetic jets after the decay ({\it i.e.}, the
gravitino LSP with bi-linear RPV), (II) the emission of only
non-monochromatic leptons ({\it i.e.}, the gravitino LSP with
tri-linear RPV), and (III) the emission of only monochromatic leptons
({\it i.e.}, the sneutrino LSP with tri-linear RPV). It should be
noted that there are other candidates for the LSP; the lightest
neutralino is a popular one. Cosmic-ray fluxes from the decay of the
lightest neutralino are similar to those obtained in the gravitino LSP
case. We thus omit to study the case of neutralino LSP.

For the BG electron flux which originates in SNR, we take $\gamma^e$
and the normalization of SNR electrons to be free parameters. We then
discuss when the decaying dark matter scenario gives consistent result
with the present cosmic-ray observations. At the beginning of this
section, we first give our numerical procedure to determine the
``best-fit'' values of the model parameters. Then, we will show the
results of our numerical analysis.

\subsection{Numerical procedure}
\label{sec:procedure}
\setcounter{equation}{0}

In order to determine the values of the model parameters preferred by
the results of PAMELA and Fermi-LAT experiments, we define the
$\chi^2_e$-variable as
\begin{eqnarray}
  \chi^2_e = \chi^2_{\rm Pamela} + \chi^2_{\rm Fermi}.
\end{eqnarray}
Here, variables $\chi^2_{\rm Pamela}$ and $\chi^2_{\rm Fermi}$ are defined as
\begin{eqnarray}
  \chi^2_{\rm Pamela} 
  =
  \sum_{i}^{N_{\rm P}} 
  \frac{\left(R^{\rm (th)}_{e^+,i}-R^{\rm (obs)}_{e^+,i}\right)^2}
  {\Delta R_{e^+,i}^2},
  \qquad
  \chi^2_{\rm Fermi}
  =
  \sum_{i}^{N_{\rm F}} 
  \frac{\left(\Phi^{\rm (th)}_{e^++e^-,i}-\Phi^{\rm (obs)}_{e^++e^-,i}\right)^2}
  {\Delta \Phi_{e^++e^-,i}^2},
\end{eqnarray}
where $R^{\rm (th)}_{e^+,i}$ ($R^{\rm (obs)}_{e^+,i}$) is simulated
(observed) positron fraction in the $i$-th bin, and $\Phi^{\rm
  (th)}_{e^++e^-,i}$ ($\Phi^{\rm (obs)}_{e^++e^-,i}$) is simulated
(observed) total ($e^++e^-$) flux. $\Delta R_{e^+,i}$ and $\Delta
\Phi_{e^++e^-,i}$ are observational errors in the $i$-th bin, and
$N_{\rm P}$ and $N_{\rm F}$ are the number of data points for
those. Since effects of the solar modulation to cosmic-ray $e^{\pm}$
is conspicuous in the energy range below $O(10~{\rm GeV})$, we use
five data points of the PAMELA experiment above 15 GeV in the
calculation of $\chi^2_{\rm Pamela}$.  Then, $N_{\rm P}=5$ (whereas
$N_{\rm F}=26$).  In addition, we neglect the systematic error from
the energy calibration in the Fermi-LAT data unless otherwise
mentioned; the effect of the systematic error is not so important for
most of the cases.

We calculate the $\chi^2_e$-variable defined above as a function of
$m_{\rm DM}$, $\tau_{\rm DM}$, $\gamma^e$, and $a_e$. Because our
primary purpose is to find a solution to the PAMELA anomaly, we first
calculate $\chi^2_{\rm Pamela}$ with $m_{\rm DM}$ being fixed, and
identify 95\% C.L. allowed region (corresponding to $\chi^2_{\rm
  Pamela}<11.07$) on the ($\tau_{\rm DM}$, $\gamma^e$,
$a_e$)-space. After that, in the parameter region allowed by the
PAMELA data, we find out parameter space which satisfies
$\chi^2_e<44.99$, which corresponds to the 95\% C.L. allowed region
(for 31 degrees of freedom). We also search the point where the total
$\chi^2_e$ is minimized; we call such a point as the ``best-fit''
point. Using the parameters, we also simulate cosmic $\gamma$-ray and
cosmic-ray $p$ and $\bar{p}$ to check consistency with those latest
observations.

\subsection{Gravitino dark matter with bi-linear RPV}
\label{subsec:bi-linear}

\begin{figure}
\begin{center}
\includegraphics[scale=0.95]{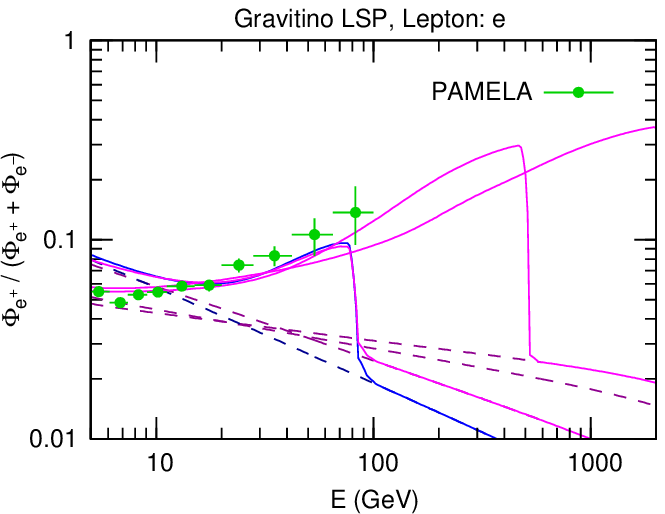}
\qquad
\includegraphics[scale=0.95]{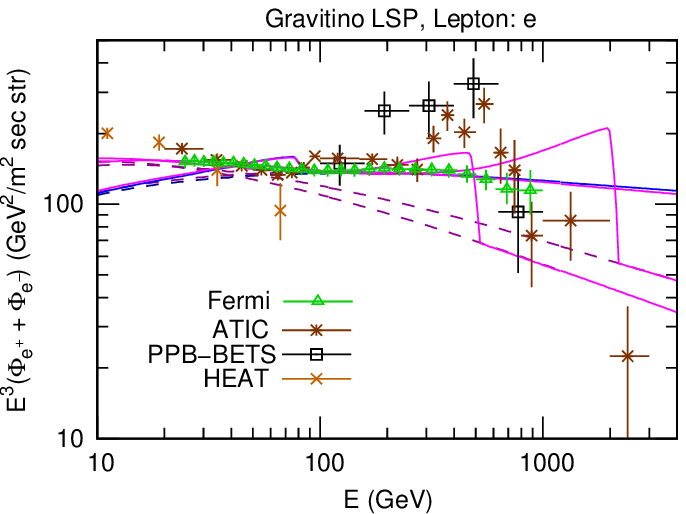}
\\
\vspace{0.5cm}
\includegraphics[scale=0.95]{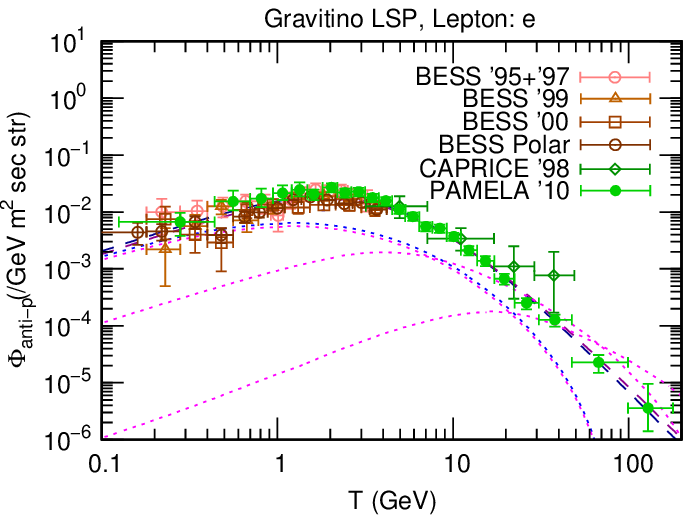}
\qquad
\includegraphics[scale=0.95]{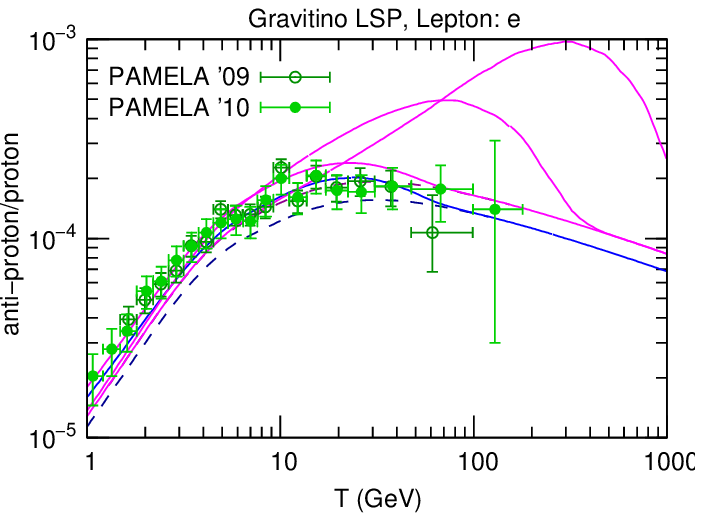}
\\
\vspace{0.5cm}
\includegraphics[scale=0.95]{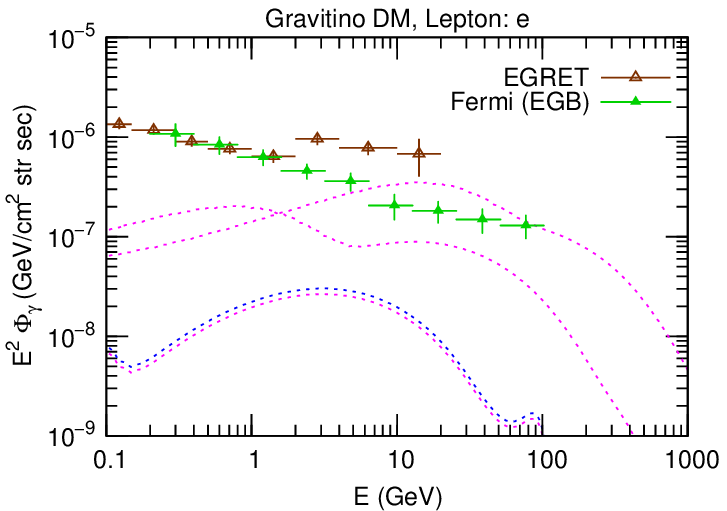} 
~~~
\includegraphics[scale=0.95]{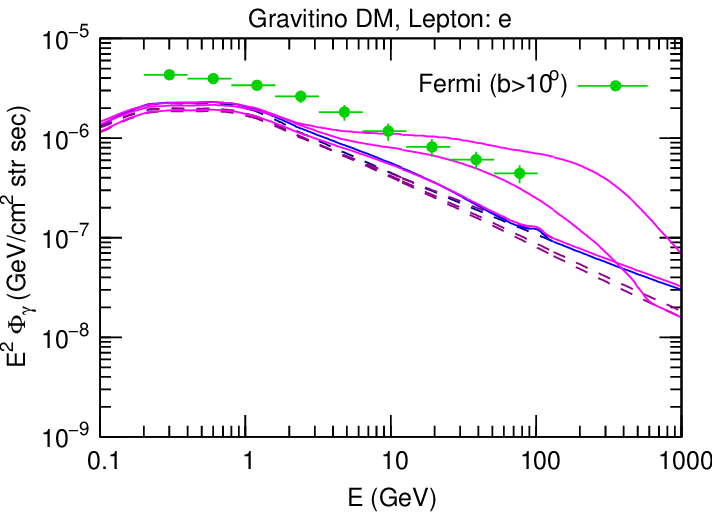} 
\caption{\small Cosmic-rays from the decaying gravitino LSP in cases
  (I-1) to (I-4) in Table \ref{table:Summary}, where positron fraction
  (top left), total $(e^++e^-)$ flux (top right), anti-proton flux
  (middle left), anti-proton to proton ratio (middle right),
  extra-Galactic $\gamma$-ray flux (bottom left), and the flux from
  inside and outside of the Galaxy (bottom right) are shown. In each
  panel, results with $\gamma^p = 2.42$ (corresponding to case (I-1),
  (I-2), and (I-3), where gravitino mass is taken as $m_{3/2}=200~{\rm
    GeV},~1~{\rm TeV}$, and 4 TeV, respectively) are shown in magenta
  lines from left to right, whereas blue line is used for those with
  $\gamma^p=2.52$ (corresponding to the case (I-4) where
  $m_{3/2}=200~{\rm GeV}$ is chosen).  Signal from the decaying dark
  matter and astrophysical BG are depicted by dotted and dashed lines,
  respectively, while solid lines correspond to signal + BG.}
\label{fig:CaseIa}
\end{center}
\end{figure}

To begin with, we study the case where the gravitino dark matter
decays dominantly into the first generation lepton through the
bi-linear RPV operator, which is the case when $B_1\gg B_2$,
$B_3$. Using the $\chi^2_e$-variable, we have found the 95\% allowed
region at $m_{3/2} \sim 200$ GeV and $m_{3/2} \gtrsim 1$ TeV. The
best-fit parameters in this case are summarized in
Table~\ref{table:Summary} as cases (I-1) to (I-3), where the values of
$\chi^2_{\rm Pamela}$ and $\chi^2_e$ in each case are also
shown. Positron fraction and the total $(e^+ + e^-)$ flux are depicted
in Fig.~\ref{fig:CaseIa} (top two panels), where observational data
are also given in each panel; the PAMELA data~\cite{Adriani:2008zr} on
the top left panel, while Fermi-LAT~\cite{Abdo:2009zk},
ATIC~\cite{Chang:2008zzr}, PPB-BETS~\cite{Torii:2008xu}, and HEAT
data~\cite{DuVernois:2001bb} on the top right one.

Our result shows that, if $m_{3/2}\sim 200\ {\rm GeV}$, the decaying
gravitino LSP may explain the PAMELA data with being consistent with
the Fermi-LAT observation. In the present case, a monochromatic
$e^{\pm}$ is emitted in the decay, resulting in a sharp edge in the
positron fraction at $E \sim$ 100 GeV. On the other hand, in the total
flux, $e^{\pm}$ from dark matter is overwhelmed by BG flux, and hence
the edge in the total flux is almost hidden by BG.  Even though a
small edge is visible in the total flux, the value of $\chi^2_e$
indicates that the model is allowed at 95\ \% C.L..  (See Table
\ref{table:Summary}.)  In addition, we have also checked that the
$\chi^2_{\rm Fermi}$ variable calculated solely from the total flux is
smaller than the 95\ \% C.L. bound.\footnote
{It is also described in \cite{Hooper:2009cs} that the scenario which
  gives such a sharp edge in the flux is not excluded by $\chi^2$
  analysis.  }
Thus, we conclude that reasonable
agreements with both the PAMELA and Fermi-LAT data are realized when
$m_{3/2}\sim 200~{\rm GeV}$. Such a possibility has not been
explicitly described in previous works after the Fermi-LAT ($e^++e^-$)
data was released.\footnote{This possibility in the framework of the
  decaying dark matter with a few hundred GeV mass was pointed out in
  the scenario where the dark matter decays dominantly into
  $\mu^+\mu^-$ pair~\cite{Barger:2009yt}.} On the other hand, with
larger gravitino mass, agreement with the PAMELA data becomes good but
not with the Fermi-LAT data.

Concerning cosmic-ray proton and anti-proton, the $\bar{p}$ flux is
shown in the middle left panel of Fig.~\ref{fig:CaseIa}. Here we take
the same model parameters as in top two panels. Observational data by
BESS~\cite{Orito:1999re}, CAPRICE~\cite{Boezio:2001ac} and
PAMELA~\cite{:2010rc} experiments are also shown in the panel. It can
be seen that the signal from the decay of dark matter is smaller than
background in most of the energy range, irrespective of the gravitino
mass.\footnote{Some people may think that the result seems to be
  inconsistent with the previous understanding about $\bar{p}$ from
  the hadronic decay. In fact, the flux simulated in the transport
  equation within the GALPROP code gives that between MED and MIN
  models proposed in Ref.~\cite{Donato:2003xg}.} However, if we
compare the simulated result with the latest observation of the ratio
between anti-proton and proton ($\bar{p}$/$p$) by the PAMELA
experiment, the scenario turns out to be strongly constrained. On the
middle right panel in Fig.~\ref{fig:CaseIa}, we plot the ratio, where
observational data of the PAMELA experiment~\cite{:2010rc,
  Adriani:2008zq} is also shown. From the figure, it can be seen that
the ratio becomes significantly larger than the observed one if the
gravitino mass is large. Even in the case of $m_{3/2} \sim 200\ {\rm
  GeV}$, the predicted ratio is slightly larger than the observation.

Here, we should recall that the BG proton flux depends on injection
index and normalization of SNR nucleon spectrum, whose values are
determined by fitting the observed flux of proton, B/C ratio, and so
on. By taking account of the uncertainties of these observations, one
can vary the values of injection index and normalization these
parameters, which may relax the severe constraint. Let us take
$\gamma^p =$ 2.52 instead of $\gamma^p =$ 2.42 above 4 GeV with the
normalization being unchanged from the value adopted in the
Conventional model.\footnote{By the explicit simulation of proton and
  B/C ratio, we have checked that their spectra are not affected by
  this change over 10 GeV. Less than 10 GeV, the flux is significantly
  affected by solar modulation. We thus ignore the flux under such low
  energy. Also, we have checked that the spectrum of $\gamma$-ray from
  pion decay is almost unchanged. See later discussion and Appendix.}
The result with this choice is summarized in Table~\ref{table:Summary}
as the case (I-4), and $\bar{p}$ flux and $\bar{p}/p$ ratio are
depicted in middle two panels of Fig.\ref{fig:CaseIa} (blue line).  We
calculated $\chi^2$ variable based on the $\bar{p}/p$ ratio given in
the latest PAMELA data \cite{:2010rc}.  Then, we found
$\chi_{\bar{p}/p}^2=12.3$ (with the degree of freedom being 23), and
hence the $\bar{p}/p$ ratio becomes consistent with the PAMELA
observation at 95 \% C.L.  We also reevaluate positron fraction and
total ($e^+ + e^-$) flux taking $\gamma^p =$ 2.52 above 4 GeV.  The
results are also shown in top two panels in Fig.~\ref{fig:CaseIa}
(blue lines). One can see that the fluxes of $e^\pm$ are almost
unchanged. Therefore, $p$ and $\bar{p}$ fluxes in the decaying
gravitino scenario can be in a reasonable agreement with the PAMELA
data if $m_{3/2} \sim$ 200 GeV. On the contrary, when the gravitino
mass is larger, the spectrum of $\bar{p}/p$ becomes inconsistent with
the observation of PAMELA.

We also simulate $\gamma$-ray flux, since it is inevitably produced
from cosmic-ray $e^{\pm}$ via IC and bremsstrahlung processes inside
Galaxy as well as via IC process in extra-Galactic
region. Furthermore, in the present case, $\gamma$-rays are also
directly produced as a consequence of gravitino decay and following
cascade decays. In the bottom left panel in Fig.~\ref{fig:CaseIa},
$\gamma$-ray from extra-Galactic region due to the decaying gravitino
is given. Again, we take model parameters as in top two
panels. Extra-Galactic $\gamma$-ray background (EGB) data given by
EGRET~\cite{Sreekumar:1997un} and Fermi-LAT~\cite{Abdo:2010nz}
experiments are also shown. One can see that $\gamma$-ray flux from
the decaying dark matter is expected to be smaller than the observed
value when $m_{3/2} \lesssim$ a few TeV. On the other hand, the flux
becomes too large to be consistent with the observation in high energy
region $E_{\gamma} >$ 10 GeV when $m_{3/2} \sim$ 4 TeV; in such an
energy region, the $\gamma$-ray is mainly from the IC process of
$e^\pm$ produced by the decay of dark matter. The gravitino LSP in
this scenario with $m_{3/2} \gtrsim$ 4 TeV is thus disfavored.

In addition to extra-Galactic $\gamma$-rays, we have simulated the
contribution from the decaying gravitino LSP inside the Galaxy. Here,
we have averaged IC-induced, bremsstrahlung-induced, and primary
$\gamma$-ray in the region $b > 10^{\circ}$ with $b$ being Galactic
latitude, and added them to the flux from extra-Galactic region to
obtain the total flux from the gravitino LSP. (We have checked that
the contribution from the bremsstrahlung process is subdominant.) The
result is shown on the bottom right panel in Fig.~\ref{fig:CaseIa}
with the observational data of the Fermi-LAT experiment
\cite{Abdo:2010nz}. From the figure, one can see that the calculation
leads to almost the same conclusion as that from the extra-Galactic
$\gamma$-ray; both primary and IC-induced one are expected to be
smaller than the observation as far as $m_{3/2} \lesssim$ a few
TeV. Therefore, by taking into account the constraints from
anti-proton and $\gamma$-ray, we can conclude that the PAMELA anomaly
can be solved with the unstable gravitino LSP with $m_{3/2} \sim 200
~{\rm GeV}$ without conflicting with other observations.

\begin{figure}
\begin{center}
\includegraphics[scale=0.95]{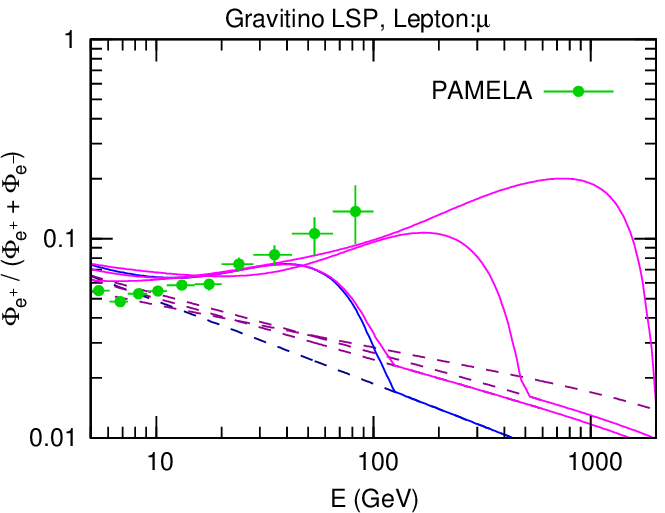}
\qquad
\includegraphics[scale=0.95]{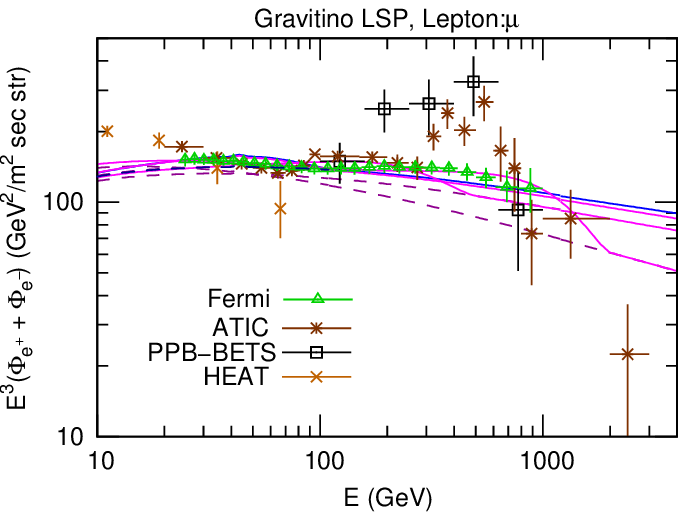}
\\
\vspace{0.5cm}
\includegraphics[scale=0.95]{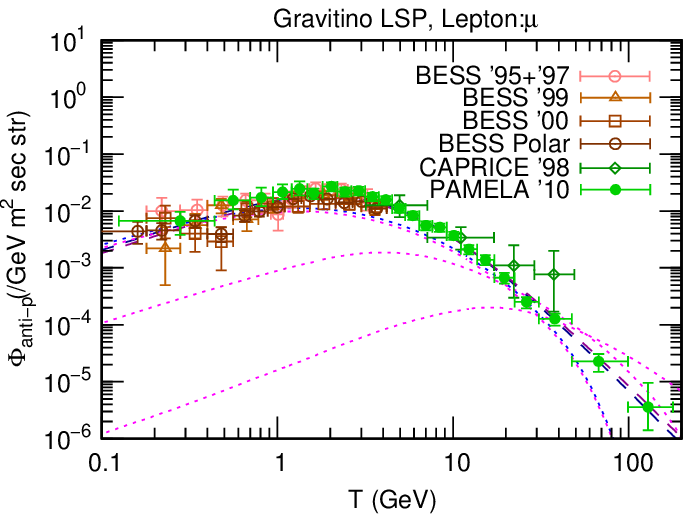}
\qquad
\includegraphics[scale=0.95]{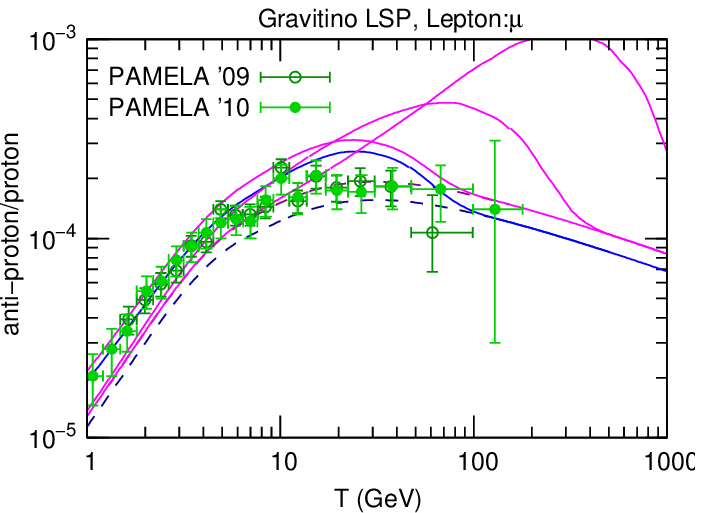}
\\
\vspace{0.5cm}
\includegraphics[scale=0.95]{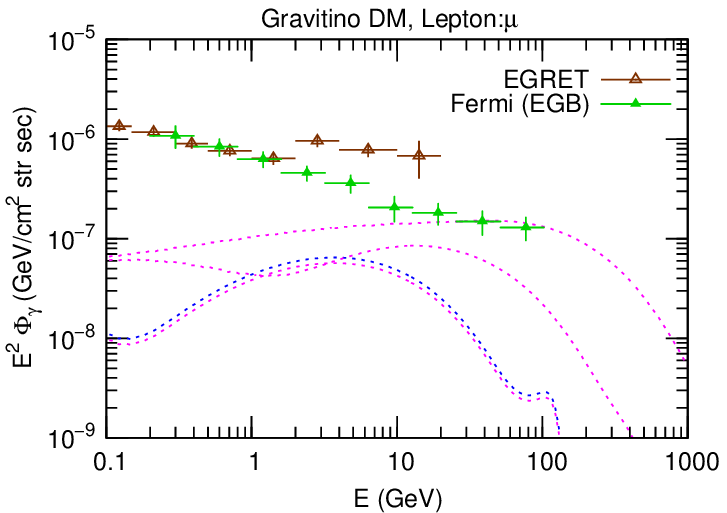}
~~~
\includegraphics[scale=0.95]{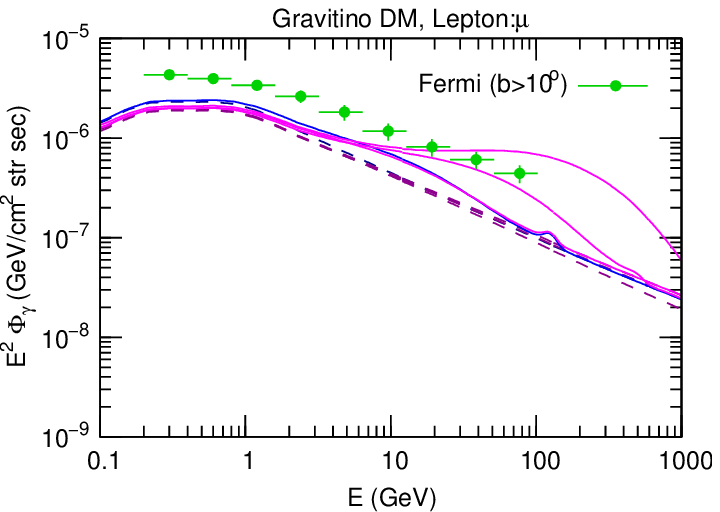}
\caption{\small Cosmic-rays from the decaying gravitino LSP in cases
  (I-5) $-$ (I-8) in Table \ref{table:Summary}.  Each panel shows
    cosmic-ray as the same way as in Fig.~\ref{fig:CaseIa}.
    Results with $\gamma^p = 2.42$ (corresponding to case (I-5),
    (I-6), and (I-7), where gravitino mass is taken as
    $m_{3/2}=250~{\rm GeV},~1~{\rm TeV}$, and 4 TeV, respectively) in
    magenta from left to right, and one with $\gamma^p=2.52$
    (corresponding to the case (I-8) where the mass is taken to be 250
    GeV) in blue are shown.  }
  \label{fig:CaseIb}
\end{center}
\end{figure}

Next, we consider the case where the gravitino LSP decays into the
second generation lepton. We have found the allowed region of
$200~{\rm GeV} \lesssim m_{3/2} \lesssim 500~{\rm GeV}$ and
$m_{3/2}\gtrsim 1~{\rm TeV}$. Model parameters giving the best-fit
results to cosmic-ray observations are summarized in
Table~\ref{table:Summary} as the cases (I-5) to (I-8). The predicted
cosmic-ray fluxes using the best-fit parameters are shown in
Fig.~\ref{fig:CaseIb}. It can be seen that the positron fraction as
well as the total $(e^+ + e^-)$ flux well agree with the results of
PAMELA and Fermi-LAT experiments irrespective of the gravitino mass
(top two panels). It is also found that the gravitino LSP with its
mass over 1 TeV results in a better agreement with the Fermi-LAT data
than the previous case.

As in the previous case, the observations of anti-proton flux give
stringent constraint on the gravitino mass. In middle two panels in
Fig.~\ref{fig:CaseIb}, we show the anti-proton flux and $\bar{p}/p$
ratio. The numerical result shows that large gravitino mass is
disfavored from the observation of $\bar{p}/p$ at the PAMELA
experiment. However, when the gravitino mass is small, the simulated
$\bar{p}/p$ is consistent with the latest observation at 95 \%
C.L.. (For $m_{3/2}=250\ {\rm GeV}$, for example, we found
$\chi_{\bar{p}/p}^2=31.3$.) We also plot $\gamma$-rays in bottom two
panels in Fig.~\ref{fig:CaseIb}, from which we see that the
$\gamma$-ray flux becomes comparable to the observation when $m_{3/2}
\sim$ 4 TeV. This $\gamma$-ray is mainly the primary one directly
produced by the decay of the gravitino LSP. The IC-induced
$\gamma$-ray is not significant, because the primary $e^{\pm}$
produced by the gravitino LSP is softer than the previous case, so
that the IC-induced $\gamma$-ray is suppressed. Therefore, we conclude
that PAMELA anomaly can be solved in the present scenario for
$200~{\rm GeV} \lesssim m_{3/2}\lesssim 500~{\rm GeV}$.

\subsection{Gravitino dark matter with tri-linear RPV}
\label{subsec:tri-linear}

\begin{figure}
\begin{center}
\includegraphics[scale=0.95]{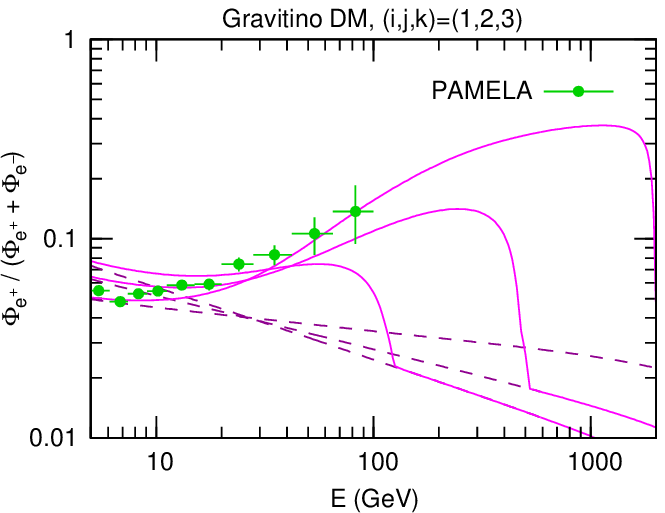}
\qquad
\includegraphics[scale=0.95]{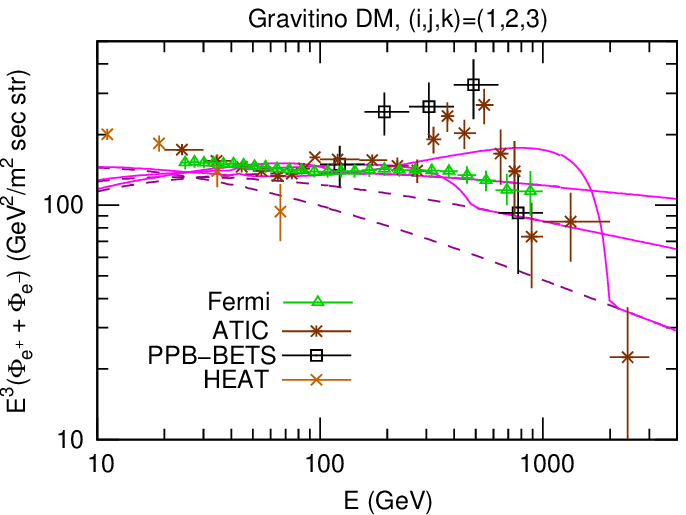}
\\
\vspace{0.5cm}
\includegraphics[scale=0.95]{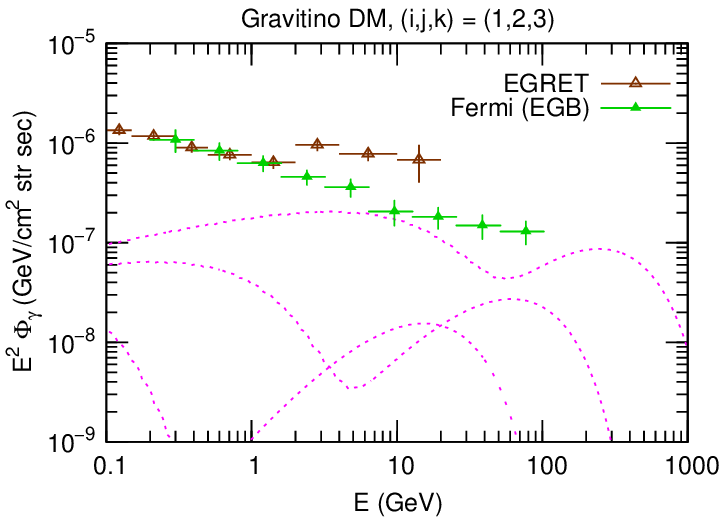}
~~~
\includegraphics[scale=0.95]{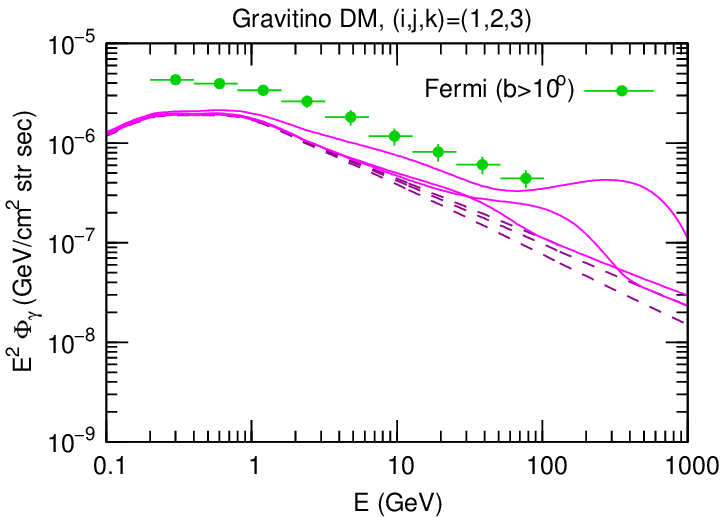}
\caption{\small Cosmic-rays from the decaying gravitino LSP in the
  cases (II-1) to (II-3) in Table \ref{table:Summary}, where positron
  fraction (upper left), total $(e^+ + e^-)$ flux (upper light),
  extra-Galactic $\gamma$-ray flux (lower left), and the flux from
  inside and outside of the Galaxy (lower right) are shown.  Line type
  is assigned as the same as in Fig.~\ref{fig:CaseIa}, {\it i.e.},
  signal from the decaying dark matter and astrophysical BG are
  depicted by dotted and dashed lines, respectively, while solid lines
  correspond to signal + BG. In each panel, numerical results
  correspond to (II-1), (II-2), and (II-3) from left to right (bottom
  to top in lower left panel), where gravitino mass are taken as
  $m_{3/2}=250~{\rm GeV}$, 1 TeV, and 4 TeV, respectively.}
\label{fig:CaseII}
\end{center}
\end{figure}

When the gravitino LSP decays mainly through the tri-linear RPV
operator given in Eq.~\eqref{eq:W_LLE}, the final state of the decay
is composed of only (three) leptons. In such a case, constraints from
anti-proton flux as well as the ratio between anti-proton and proton
are irrelevant in the study of the decaying dark matter scenario.

With the $\chi^2_e$-analysis, we have found the region consistent with
$e^\pm$ observations at 95\% C.L. in 200 GeV $\lesssim m_{3/2}
\lesssim$ 400 GeV and 1 TeV $\lesssim m_{3/2} \lesssim$ 3 TeV. (In the
numerical calculation, we take $m_{\tilde{l}_R}=1.2 ~m_{3/2}$.) The
model parameters for the best-fit results are again summarized in
Table~\ref{table:Summary} as cases (II-1) to (II-3), and simulated
results for cosmic-ray $e^\pm$ and $\gamma$-rays are shown in
Fig.~\ref{fig:CaseII}. In upper two panels, positron fraction and
total $(e^++e^-)$ flux are shown. Here, we consider the case where the
component $\lambda_{123}$ is dominant compared to others. It can be
seen that the fitting to the Fermi-LAT data becomes worse in the large
$m_{3/2}$ region though it is possible to explain the PAMELA data
well.

We have calculated $\gamma$-ray flux. Here, we consider the case in
which $\lambda_{123}$ is the only relevant RPV parameter, so we also
calculate primary $\gamma$-ray flux from the decay of $\tau$, along
with the one from the IC process. The results are shown in lower two
panels in Fig.~\ref{fig:CaseII}; the left one is extra-Galactic
contribution and the right one is extra- plus inner-Galactic
contribution. For the extra-Galactic one, the simulated flux is much
smaller than the observation unless $m_{3/2} \gtrsim 4~{\rm TeV}$.
The observation of total $\gamma$-ray (for $b>10^{\circ}$) gives
almost the same upper bound on the mass.  As a consequence, the
parameter region favored by the observation of $e^{\pm}$ ({\it i.e.},
200 GeV $\lesssim m_{3/2} \lesssim$ 400 GeV and 1 TeV $\lesssim
m_{3/2} \lesssim$ 3 TeV) are not excluded by the $\gamma$-ray
observation at the Fermi-LAT experiment. It can be also seen that the
total $\gamma$-ray flux at high energy region is noticeable when
$m_{3/2}\gtrsim 4~{\rm TeV}$. This intense flux comes from the decay
of $\tau$.

In addition to the case where the component $\lambda_{123}$ dominates,
we have also considered the case in which only $\lambda_{121}$ or
$\lambda_{122}$ is relevant.\footnote
{In those cases, there is no $\gamma$-ray primarily produced by the
  decay.  }
In the former case, however, injected $e^{\pm}$ is too hard to be
consistent with observations at 95\% C.L. in the entire range of
$m_{3/2}$. On the other hand, in the latter case, the same conclusion
as in the case $\lambda_{123}$ dominates holds except that the allowed
region at the TeV scale disappears because the predicted total $(e^+ +
e^-)$ flux hardly agrees with the Fermi result.

\subsection{Sneutrino dark matter with tri-linear RPV}
\label{subsec:snutri-linear}

\begin{figure}
\begin{center}
\includegraphics[scale=0.95]{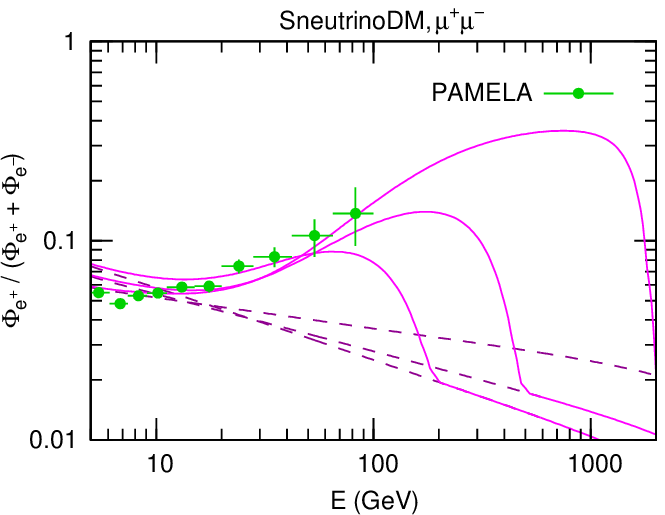}
\qquad
\includegraphics[scale=0.95]{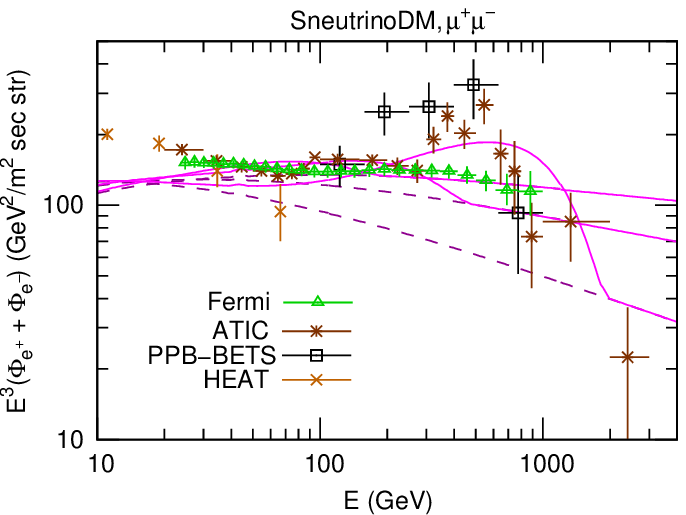}
\\
\vspace{0.5cm}
\includegraphics[scale=0.95]{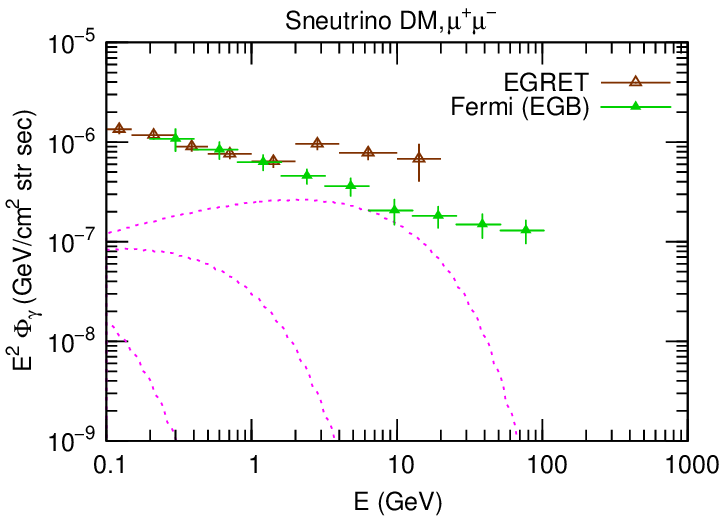}
~~~
\includegraphics[scale=0.95]{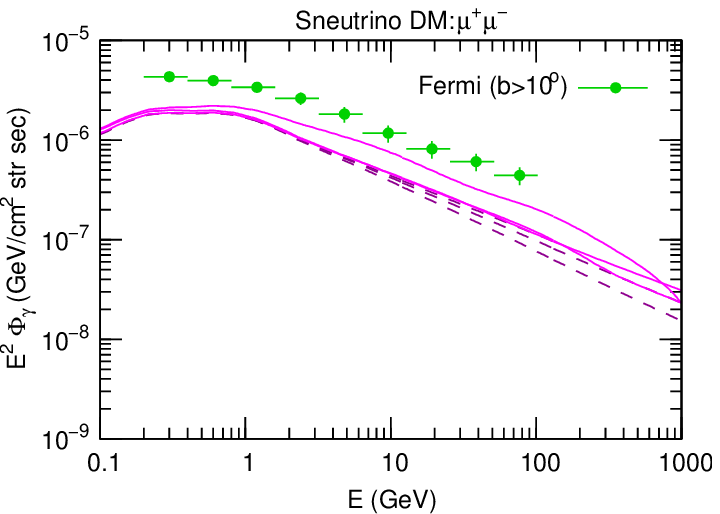}
\caption{\small Cosmic-rays from the decaying sneutrino LSP in the
  cases (III-1) to (III-3) in Table \ref{table:Summary}.  The
    panel position and line contents are the same as in
    Fig.~\ref{fig:CaseII}. In each panel, lines from left to right
    (bottom to top in lower left panel) correspond to the case
    (III-1), (III-2), and (III-3), where sneutrino mass is taken as
    $m_{\tilde{\nu}}=400~{\rm GeV}, 1~{\rm TeV}$, and 4 TeV,
    respectively. }
\label{fig:CaseIII}
\end{center}
\end{figure}

So far, we have discussed the gravitino LSP cases. However, there are
other candidates for LSP. In particular, as we have mentioned, the
sneutrino LSP may provide decay modes different from the gravitino
LSP. In particular, with the superpotential given in Eq.\
\eqref{eq:W_LLE}, the sneutrino LSP may decay into two charged
leptons. We then expect different cosmic-ray spectra from those of the
gravitino LSP. In this subsection, we briefly comment on the behaviors
of cosmic-ray fluxes in the sneutrino dark matter scenario with
tri-linear RPV.  The important effects are on $e^\pm$ and $\gamma$-ray
fluxes, but not on anti-proton flux.

In Fig.\ \ref{fig:CaseIII}, we show the positron fraction (upper left)
and the total $(e^+ + e^-)$ flux (upper right) for the case where the
sneutrino dominantly decays through the process
$\tilde{\nu}\rightarrow \mu^+\mu^-$. The parameters to give those
results are summarized in Table~\ref{table:Summary} as cases (III-1)
to (III-3). 

With the use of the $\chi^2_e$-analysis, we have found 95\%
C.L. allowed region in $300~{\rm GeV} \lesssim m_{\tilde{\nu}}
\lesssim 500~{\rm GeV}$, with neglecting the possible systematic error
from the energy calibration in the Fermi-LAT data.  For
$m_{\tilde{\nu}} \gtrsim 500~{\rm GeV}$, the total flux becomes
inconsistent with Fermi-LAT data, while positron fraction still agrees
with the PAMELA data.  If we take account of the effect of systematic
error in Fermi-LAT data, the allowed region may become larger; in
particular, adopting the $10\ \%$ reduction of the energy, the region
$1~{\rm TeV}\lesssim m_{\tilde{\nu}}\lesssim 3~{\rm TeV}$ becomes
allowed at 95\% C.L.  This result is consistent with
\cite{Barger:2009yt}.

We also give $\gamma$-ray flux in Fig.~\ref{fig:CaseIII}. The only
relevant process to produce $\gamma$-ray here is the IC
scattering.\footnote
{There could exist contribution from final state radiation (FSR).  FSR
  may give hard $\gamma$-ray and also give important contribution when
  the leptonic decay is suppressed by kinematics or chirality; however
  it is not such a case which we consider. }
In the figure, IC-induced $\gamma$-ray from extra-Galactic region
(inner-plus extra-Galactic region) is given lower left (right)
panel. Although the flux is less than the observed data in
$b>10^{\circ}$, IC-induced $\gamma$-ray in the extra-Galactic region
becomes comparable to or larger than the data when
$m_{\tilde{\nu}}\gtrsim 4~{\rm TeV}$. This is the same result obtained
in the previous three-body decay case. We thus conclude that the
parameter region $300~{\rm GeV}\lesssim m_{\tilde{\nu}} \lesssim
500~{\rm GeV}$ gives good fit with PAMELA and Fermi-LAT ($e^++e^-$)
data and the region is not constrained by $\gamma$-ray observation.

We have also considered the decaying scenario of
$\tilde{\nu}\rightarrow e^+e^-$. In such a case, however, we could not
find the allowed region. This is because the monochromatic electron
and positron gives very sharp edge in the flux so that the flux does
not agree with the data of $(e^++e^-)$ flux though the fit with the
PAMELA data is good.

\section{Conclusion}
\label{sec:conclusion}
\setcounter{equation}{0}

In this article, we have calculated cosmic-ray fluxes from decaying
dark matter as well as background in the same propagation model by
using GALPROP. Aiming for explaining the PAMELA anomaly with being
consistent with other cosmic-ray observations, we have reevaluated the
background cosmic-ray fluxes.  Cosmic-rays from decaying dark matter,
on the other hand, is strongly dependent on the spectra of final-state
particles. If one specifies the distributions of final-state
particles, the cosmic-ray spectra are determined independently of
detailed framework of the model of decaying dark matter.

To make our discussion concrete, we have studied gravitino dark matter
in $R$-parity violated supersymmetric model. Under $R$-parity
violation, gravitino dominantly decays to $Wl_i$ in bi-linear RPV,
while it decays to $\nu_i l^+_k l^-_j$ in tri-linear one. In the
former scenario, we have found that the simulated cosmic-ray $e^{\pm}$
flux from dark matter and background agrees with the PAMELA and
Fermi-LAT data, irrespective of gravitino mass. However, it has been
shown that the production of cosmic-ray $\bar{p}$ becomes enhanced
when the mass is large; thus the gravitino mass larger than $\sim 300
~{\rm GeV}$ is disfavored. For $\gamma$-ray, the flux is consistent
with the observation as far as $m_{3/2}\lesssim 4~{\rm TeV}$. In the
latter case, it has been found that the PAMELA anomaly can be
explained in the mass region $200~{\rm GeV} \lesssim m_{3/2} \lesssim
400~{\rm GeV}$ or $1~{\rm TeV} \lesssim m_{3/2} \lesssim 3~{\rm TeV}$,
being consistent with Fermi-LAT data. When the mass is larger than
$\sim 4~{\rm TeV}$, the fit with each data becomes worse. In addition,
IC-induced $\gamma$-ray constrain such large mass region.

We have also considered the case where sneutrino is the LSP, assuming
that it decays into lepton pair via tri-linear RPV interaction. We
have seen that, when the dominant decay mode is
$\tilde{\nu}\rightarrow\mu^+\mu^-$, the positron fraction can be in a
good agreement with the PAMELA data without conflicting the Fermi
result when $300~{\rm GeV}\lesssim m_{\tilde{\nu}} \lesssim 500~{\rm
  GeV}$. With such a choice of the sneutrino mass, $\gamma$-ray flux
induced by the dark matter decay is much smaller than the observed
one. In addition, in this case, it should be noted that the constraint
from anti-proton flux is irrelevant because hadrons are hardly
produced by the decay of $\tilde{\nu}$.\\

\vspace{0.5cm}
\noindent
{\bf Acknowledgments}\\

\vspace{-0.3cm}

\noindent
This work was supported in part by Research Fellowships of the Japan
Society for the Promotion of Science for Young Scientists (K.I.), and
by the Grant-in-Aid for Scientific Research from the Ministry of
Education, Science, Sports, and Culture of Japan, No.\ 21740174 (S.M.)
and Nos.\ 22540263 and 22244021 (T.M.). We also thank organizers,
especially, at Niigata University of SI2010(YITP-W-10-07) in
Fuji-Yoshida, and the Yukawa Institute for Theoretical Physics at
Kyoto University. Discussions during the workshop were useful to
finalize this project (S.M. and T.M.).

\appendix
\section{Background Cosmic-Rays}
\label{appendix}
\setcounter{equation}{0}

\begin{figure}
  \begin{center}
  \includegraphics[scale=0.95]{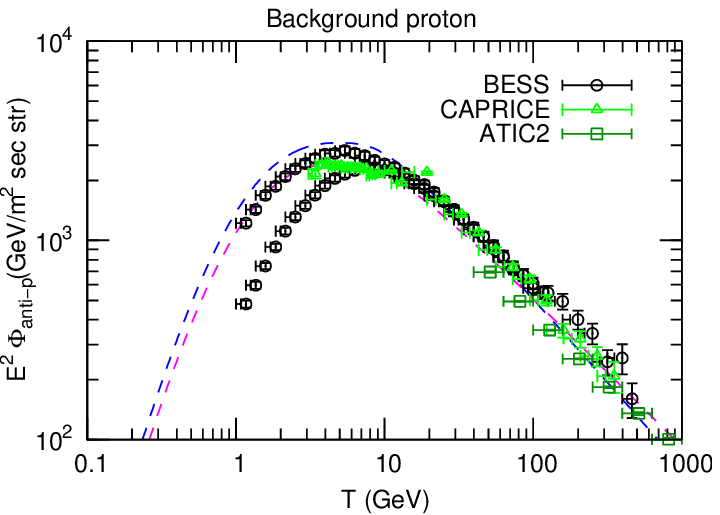} 
  \qquad
  \includegraphics[scale=0.95]{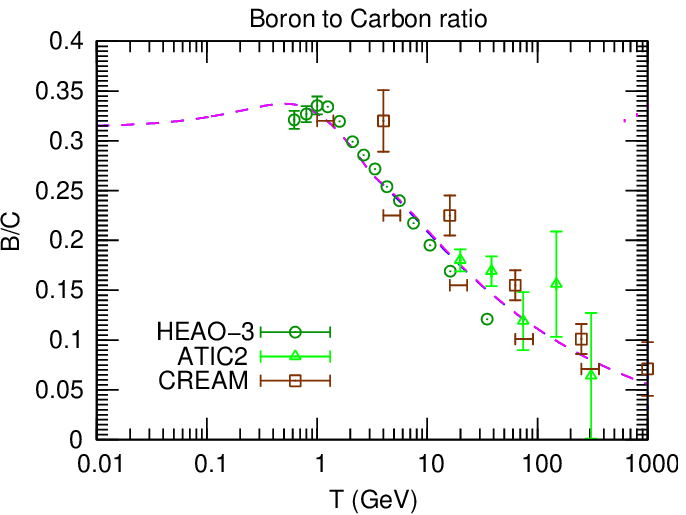}
  \caption{\small Background proton flux (the left panel) and the B/C
    ratio (the right panel). Here, the blue (magenta) line shows the
    simulated result by $\gamma^p = 2.52$ (2.42) above 4 GeV. Other
    parameters are taken as the same in Conventional model.}
  \end{center}
  \label{fig:BGnuclei}
\end{figure}

We consider background cosmic-rays predicted in the Conventional
model, except for taking $\gamma^p = 2.52$ (instead of 2.42) above 4
GeV. The main purpose in this appendix is to show that the change of
$\gamma^p$ does not significantly affect background cosmic-ray fluxes,
and that such a choice is consistent with observations. Cosmic-ray $p$
flux and the B/C ratio are shown in Fig.~5, where
simulated results are depicted with a blue line. The results with
$\gamma^p=2.42$ above 4 GeV are also shown for comparison with a
magenta line. For the $p$ flux shown in the left panel of the figure,
it can be seen that the spectrum with $\gamma^p=2.52$ becomes slightly
softer than that with $\gamma^p=2.42$ as expected, and both spectra
are well consistent with observations.\footnote{In the region $E
  \lesssim$ 10 GeV, the flux with $\gamma^p = 2.52$ seems to slightly
  exceed the observations, it can be, however, optimized by choosing a
  proper normalization of the SNR $p$ flux. Moreover, in this region,
  the flux is affected by the effect of solar modulation, which leads
  to further uncertainties. }

On the other hand, the B/C ratio is expected to be hardly
changed. This is because the fluxes of B and C are mostly determined
by the secondary and primary fluxes from SNR and hence the B/C ratio
depends strongly on diffusion parameters but is insensitive to SNR
spectra. In fact, as shown in the right panel of
Fig.~\ref{fig:BGnuclei}, there is no difference between $\gamma^p =
2.42$ and 2.52 cases, which are consistent with observations. (The
intensity in $E\lesssim 1~{\rm GeV}$ is larger than the one appeared
in Ref.~\cite{Strong:2004de}. This is due to the different choice of
solar modulation potential; we take $\phi=550~{\rm MV}$, whereas
$\phi=450~{\rm MV}$ is chosen in Ref.~\cite{Strong:2004de}.)


\begin{thebibliography}{99}

\bibitem{Adriani:2008zr}
  O.~Adriani {\it et al.}  [PAMELA Collaboration],
  Nature {\bf 458}, 607 (2009).

\bibitem{Hooper:2008kg}
  D.~Hooper, P.~Blasi and P.~D.~Serpico,
  JCAP {\bf 0901}, 025 (2009).

\bibitem{Nomura:2008ru}
  Y.~Nomura and J.~Thaler,
  Phys.\ Rev.\  D {\bf 79}, 075008 (2009).
\bibitem{Yin:2008bs}
  P.~f.~Yin, Q.~Yuan, J.~Liu, J.~Zhang, X.~j.~Bi and S.~h.~Zhu,
  Phys.\ Rev.\  D {\bf 79}, 023512 (2009).
\bibitem{Bai:2008jt}
  Y.~Bai and Z.~Han,
  Phys.\ Rev.\  D {\bf 79}, 095023 (2009).

\bibitem{Chen:2008md}
  C.~R.~Chen, F.~Takahashi and T.~T.~Yanagida,
  Phys.\ Lett.\  B {\bf 673}, 255 (2009).
\bibitem{Hamaguchi:2008rv}
  K.~Hamaguchi, E.~Nakamura, S.~Shirai and T.~T.~Yanagida,
  Phys.\ Lett.\  B {\bf 674}, 299 (2009).
\bibitem{Ponton:2008zv}
  E.~Ponton and L.~Randall,
  JHEP {\bf 0904}, 080 (2009).
\bibitem{Ibarra:2008jk}
  A.~Ibarra and D.~Tran,
  JCAP {\bf 0902}, 021 (2009).
\bibitem{Chen:2008qs}
  C.~R.~Chen, M.~M.~Nojiri, F.~Takahashi and T.~T.~Yanagida,
  Prog.\ Theor.\ Phys.\  {\bf 122}, 553  (2009).
\bibitem{Arvanitaki:2008hq} 
  A.~Arvanitaki, S.~Dimopoulos, S.~Dubovsky,
  P.~W.~Graham, R.~Harnik and S.~Rajendran,
  Phys.\ Rev.\  D {\bf 79}, 105022 (2009).
\bibitem{Hamaguchi:2008ta}
  K.~Hamaguchi, S.~Shirai and T.~T.~Yanagida,
  Phys.\ Lett.\  B {\bf 673}, 247 (2009).
\bibitem{Gogoladze:2009kv}
  I.~Gogoladze, R.~Khalid, Q.~Shafi and H.~Yuksel,
  Phys.\ Rev.\  D {\bf 79}, 055019 (2009).
\bibitem{Hamaguchi:2009sz}
  K.~Hamaguchi, F.~Takahashi and T.~T.~Yanagida,
  Phys.\ Lett.\  B {\bf 677}, 59 (2009).
\bibitem{Nardi:2008ix}
  E.~Nardi, F.~Sannino and A.~Strumia,
  JCAP {\bf 0901}, 043 (2009).
\bibitem{Chen:2008dh}
  C.~R.~Chen and F.~Takahashi,
  JCAP {\bf 0902}, 004 (2009).
\bibitem{Sierra:2009zq}
  D.~Aristizabal Sierra, D.~Restrepo and O.~Zapata,
  Phys.\ Rev.\  D {\bf 80}, 055010 (2009).
\bibitem{Demir:2009kc}
  D.~A.~Demir, L.~L.~Everett, M.~Frank, L.~Selbuz and I.~Turan,
  Phys.\ Rev.\  D {\bf 81}, 035019 (2010).
\bibitem{Zhang:2009ut}
  L.~Zhang, C.~Weniger, L.~Maccione, J.~Redondo and G.~Sigl,
  JCAP {\bf 1006}, 027 (2010).
\bibitem{Carone:2010ha}
  C.~D.~Carone, J.~Erlich and R.~Primulando,
  arXiv:1008.0642 [hep-ph].
\bibitem{Ishiwata:2008cv}
  K.~Ishiwata, S.~Matsumoto and T.~Moroi,
  Phys.\ Lett.\  B {\bf 675}, 446 (2009).

\bibitem{Cirelli:2008pk}
  M.~Cirelli, M.~Kadastik, M.~Raidal and A.~Strumia,
  Nucl.\ Phys.\  B {\bf 813}, 1 (2009).
\bibitem{Cholis:2008qq}
  I.~Cholis, D.~P.~Finkbeiner, L.~Goodenough and N.~Weiner,
  JCAP {\bf 0912}, 007 (2009).
\bibitem{Feldman:2008xs}
  D.~Feldman, Z.~Liu and P.~Nath,
  Phys.\ Rev.\  D {\bf 79}, 063509 (2009).
\bibitem{Fox:2008kb}
  P.~J.~Fox and E.~Poppitz,
  Phys.\ Rev.\  D {\bf 79}, 083528 (2009).
\bibitem{Bergstrom:2008gr}
  L.~Bergstrom, T.~Bringmann and J.~Edsjo,
  Phys.\ Rev.\  D {\bf 78}, 103520 (2008).
\bibitem{Barger:2008su}
  V.~Barger, W.~Y.~Keung, D.~Marfatia and G.~Shaughnessy,
  Phys.\ Lett.\  B {\bf 672}, 141 (2009).
\bibitem{Nelson:2008hj}
  A.~E.~Nelson and C.~Spitzer,
  arXiv:0810.5167 [hep-ph].
\bibitem{Harnik:2008uu}
  R.~Harnik and G.~D.~Kribs,
  Phys.\ Rev.\  D {\bf 79}, 095007 (2009).

\bibitem{DuVernois:2001bb}
  M.~A.~DuVernois {\it et al.},
  Astrophys.\ J.\  {\bf 559} 296 (2001).

\bibitem{Ishiwata:2008cu}
  K.~Ishiwata, S.~Matsumoto and T.~Moroi,
  Phys.\ Rev.\  D {\bf 78}, 063505 (2008).

\bibitem{Ibarra:2008qg}
  A.~Ibarra and D.~Tran,
  JCAP {\bf 0807}, 002 (2008).

\bibitem{Galprop}
  GALPROP Homepage, 
  {\tt http://galprop.stanford.edu/}.

\bibitem{Zhang:2008tb}
 J.~Zhang, X.~J.~Bi, J.~Liu, S.~M.~Liu, P.~F.~Yin, Q.~Yuan and S.~H.~Zhu,
 Phys.\ Rev.\  D {\bf 80}, 023007 (2009).
 \bibitem{Barger:2009yt}
  V.~Barger, Y.~Gao, W.~Y.~Keung, D.~Marfatia and G.~Shaughnessy,
  Phys.\ Lett.\  B {\bf 678}, 283 (2009).
\bibitem{Barger:2009xe}
 V.~Barger, Y.~Gao, W.~Y.~Keung and D.~Marfatia,
 Phys.\ Rev.\  D {\bf 80}, 063537 (2009).
\bibitem{Huh:2009ij}
 J.~H.~Huh and J.~E.~Kim,
 Phys.\ Rev.\  D {\bf 80}, 075012 (2009).
\bibitem{Bomark:2009zm}
 N.~E.~Bomark, S.~Lola, P.~Osland and A.~R.~Raklev,
 Phys.\ Lett.\  B {\bf 686}, 152 (2010).
\bibitem{Lin:2010fb}
 T.~Lin, D.~P.~Finkbeiner and G.~Dobler,
 arXiv:1004.0989 [astro-ph.CO].
\bibitem{Cotta:2010ej}
  R.~C.~Cotta, J.~A.~Conley, J.~S.~Gainer, J.~L.~Hewett and T.~G.~Rizzo,
  arXiv:1007.5520 [hep-ph].

\bibitem{Strong:1998fr}
  A.~W.~Strong, I.~V.~Moskalenko and O.~Reimer,
  Astrophys.\ J.\  {\bf 537}, 763 (2000)
  [Erratum-ibid.\  {\bf 541}, 1109 (2000)].

\bibitem{Strong:2004de}
  A.~W.~Strong, I.~V.~Moskalenko and O.~Reimer,
  Astrophys.\ J.\  {\bf 613}, 962 (2004).

\bibitem{Abdo:2009zk}
  A.~A.~Abdo {\it et al.}  [The Fermi-LAT Collaboration],
  Phys.\ Rev.\ Lett.\ {\bf 102}, 181101 (2009).

\bibitem{solarmd}
   L.~J.~Gleeson and W.~I.~Axford, ApJ\ 154,\ 1101 (1968).

\bibitem{Ensslin:1996ep}
  T.~A.~Ensslin, P.~L.~Biermann, P.~P.~Kronberg and X.~P.~Wu,
  Astrophys.\ J.\  {\bf 477}, 560 (1997);

\bibitem{Loeb:2000na}
  A.~Loeb and E.~Waxman,
  Nature {\bf 405}, 156 (2000);
  F.~Miniati,
  Mon.\ Not.\ Roy.\ Astron.\ Soc.\  {\bf 337}, 199 (2002).

\bibitem{Gao:1990bh}
  Y.~T.~Gao, F.~W.~Stecker, M.~Gleiser and D.~B.~Cline,
  Astrophys.\ J.\  {\bf 361}, L37 (1990);
  A.~Dolgov and J.~Silk,
  Phys.\ Rev.\  D {\bf 47}, 4244 (1993).

\bibitem{Navarro:1996gj}
  J.~F.~Navarro, C.~S.~Frenk and S.~D.~M.~White,
  Astrophys.\ J.\  {\bf 490}, 493 (1997).

\bibitem{Ishiwata:2009dk}
  K.~Ishiwata, S.~Matsumoto and T.~Moroi,
  Phys.\ Lett.\  B {\bf 679}, 1 (2009).

\bibitem{Cirelli:2009dv}
  M.~Cirelli, P.~Panci and P.~D.~Serpico,
  Nucl.\ Phys.\  B {\bf 840}, 284 (2010).

\bibitem{Hutsi:2010ai}
  G.~Hutsi, A.~Hektor and M.~Raidal,
  JCAP {\bf 1007}, 008 (2010).

\bibitem{Cline:2010ag}
  J.~M.~Cline, A.~C.~Vincent and W.~Xue,
  Phys.\ Rev.\  D {\bf 81}, 083512 (2010).

\bibitem{Abdo:2010ex}
  A.~A.~Abdo {\it et al.},
  Astrophys.\ J.\  {\bf 712}, 147 (2010).

\bibitem{Perelstein:2010fq}
  M.~Perelstein and B.~Shakya,
  arXiv:1002.4588 [astro-ph.HE].

\bibitem{Buckley:2010vg}
  M.~R.~Buckley and D.~Hooper,
  arXiv:1004.1644 [hep-ph].

\bibitem{Essig:2009jx}
  R.~Essig, N.~Sehgal and L.~E.~Strigari,
  Phys.\ Rev.\  D {\bf 80}, 023506 (2009).

\bibitem{Essig:2010em}
  R.~Essig, N.~Sehgal, L.~E.~Strigari, M.~Geha and J.~D.~Simon,
  arXiv:1007.4199 [astro-ph.CO].

\bibitem{Perelstein:2010at}
  M.~Perelstein and B.~Shakya,
  arXiv:1007.0018 [astro-ph.HE].

\bibitem{Sjostrand:2006za}
  T.~Sjostrand, S.~Mrenna and P.~Skands,
  JHEP {\bf 0605}, 026 (2006).

\bibitem{Moroi:1993mb}
  T.~Moroi, H.~Murayama and M.~Yamaguchi,
  Phys.\ Lett.\  B {\bf 303}, 289 (1993).

\bibitem{Feng:2003xh}
  J.~L.~Feng, A.~Rajaraman and F.~Takayama,
  Phys.\ Rev.\ Lett.\  {\bf 91}, 011302 (2003).

\bibitem{Ishiwata:2007bt}
  K.~Ishiwata, S.~Matsumoto and T.~Moroi,
  Phys.\ Rev.\  D {\bf 77}, 035004 (2008).

\bibitem{Asaka:2005cn}
  T.~Asaka, K.~Ishiwata and T.~Moroi,
  Phys.\ Rev.\  D {\bf 73}, 051301 (2006).

\bibitem{Campbell:1990fa}
  B.~A.~Campbell, S.~Davidson, J.~R.~Ellis and K.~A.~Olive,
  Phys.\ Lett.\  B {\bf 256}, 484 (1991);
  W.~Fischler, G.~F.~Giudice, R.~G.~Leigh and S.~Paban,
  Phys.\ Lett.\  B {\bf 258}, 45 (1991);
  H.~K.~Dreiner and G.~G.~Ross,
  Nucl.\ Phys.\  B {\bf 410}, 188 (1993).

\bibitem{Hall:1997ah}
  L.~J.~Hall, T.~Moroi and H.~Murayama,
  Phys.\ Lett.\  B {\bf 424}, 305 (1998).

\bibitem{Ishiwata:2009vx}
  K.~Ishiwata, S.~Matsumoto and T.~Moroi,
  JHEP {\bf 0905}, 110 (2009).

\bibitem{Chang:2008zzr}
  J.~Chang {\it et al.},
  Nature {\bf 456}, 362 (2008).

\bibitem{Torii:2008xu}
  S.~Torii {\it et al.},
  arXiv:0809.0760 [astro-ph].

\bibitem{Hooper:2009cs}
  D.~Hooper and K.~M.~Zurek,
  Phys.\ Lett.\  B {\bf 691}, 18 (2010).

\bibitem{Orito:1999re}
  S.~Orito {\it et al.}  [BESS Collaboration],
  Phys.\ Rev.\ Lett.\  {\bf 84}, 1078 (2000);
  Y.~Asaoka {\it et al.},
  Phys.\ Rev.\ Lett.\  {\bf 88}, 051101 (2002);
  K.~Abe {\it et al.},
  Phys.\ Lett.\  B {\bf 670}, 103 (2008).

\bibitem{Boezio:2001ac}
  M.~Boezio {\it et al.}  [WiZard/CAPRICE Collaboration],
  Astrophys.\ J.\  {\bf 561}, 787 (2001).

\bibitem{:2010rc}
  PAMELA Collaboration,
  arXiv:1007.0821 [astro-ph.HE].

\bibitem{Donato:2003xg}
  F.~Donato, N.~Fornengo, D.~Maurin and P.~Salati,
  Phys.\ Rev.\  D {\bf 69}, 063501 (2004).

\bibitem{Adriani:2008zq}
  O.~Adriani {\it et al.},
  Phys.\ Rev.\ Lett.\  {\bf 102}, 051101 (2009).

\bibitem{Sreekumar:1997un}
  P.~Sreekumar {\it et al.}  [EGRET Collaboration],
  Astrophys.\ J.\  {\bf 494}, 523 (1998).

\bibitem{Abdo:2010nz}
  A.~A.~Abdo {\it et al.}  [The Fermi-LAT collaboration],
  Phys.\ Rev.\ Lett.\  {\bf 104}, 101101 (2010).

\end{thebibliography}
\end{document}